\documentclass[lettersize,journal]{IEEEtran}
\usepackage{amsmath,amsfonts}
\usepackage{algorithmic}
\usepackage{algorithm}
\usepackage{array}
\usepackage[caption=false,font=footnotesize,labelfont=rm,textfont=rm]{subfig}
\usepackage{textcomp}
\usepackage{xcolor}
\usepackage{stfloats}
\usepackage{url}
\usepackage{verbatim}
\usepackage{graphicx}
\usepackage{tabularx}
\usepackage{ragged2e}
\usepackage{cite}
\usepackage{bm}
\usepackage{cite}
\usepackage{amssymb}
\usepackage{float} 
\usepackage{caption}
\usepackage{textcomp}
\usepackage{amsthm}
\usepackage{subfig}		
\usepackage{cuted}
\usepackage{booktabs}
\usepackage{hyperref}
\usepackage{amsmath}
\usepackage{cleveref}
\newtheorem{remark}{Remark}

\begin{document}
\title{Efficient Off-Grid Near-Field Cascade Channel Estimation for XL-IRS Systems via Tucker Decomposition}

\author{Wenzhou Cao, Yashuai~Cao, Tiejun~Lv, \IEEEmembership{Senior Member,~IEEE},~and Mugen~Peng,~\IEEEmembership{Fellow,~IEEE}
\thanks{Manuscript received 5 June 2025; revised 14 December 2025; accepted 13 February 2026. This paper was supported by the National Natural Science Foundation of China under No. 62271068. (\emph{corresponding author: Tiejun Lv}.)}
\thanks{W. Cao, T. Lv and M. Peng are with the School of Information and Communication Engineering, Beijing University of Posts and Telecommunications (BUPT), Beijing 100876, China (e-mail: \{cwzhou9511, lvtiejun, pmg\}@bupt.edu.cn).}
\thanks{Y. Cao is with the School of Intelligence Science and Technology, University of Science and Technology Beijing (USTB), Beijing 100083, China (e-mail: caoys@ustb.edu.cn).}

}

\maketitle
\begin{abstract}
Accurate cascaded channel state information is pivotal for unlocking the transformative capabilities of extremely large-scale intelligent reflecting surfaces (XL-IRS) in next-generation wireless networks. However, the expansive aperture of XL-IRSs induces spherical wavefront propagation due to near-field (NF) effects, complicating cascaded channel estimation. Conventional dictionary-based methods suffer from cumulative quantization errors and prohibitive computational complexity, particularly in practical uniform planar array (UPA)-structured array systems. To overcome these issues, we firstly propose a tensor modelization method for NF cascaded channels by exploiting the tensor product among the horizontal and vertical response vectors of the UPA-structured base station (BS) and the incident-reflective array response vector of the IRS. This tensor structure leverages the spatial characteristics of cascade channel, enabling independent estimation of the factor matrices to improve channel estimation efficiency. Meanwhile, to  avoid quantization errors, we propose an off-grid cascaded channel estimation framework based on sparse Tucker decomposition. Specifically, we model the received signal as a Tucker tensor, where the sparse core tensor captures path gain-delay terms and three factor dictionary matrices are spanned by the BS  array responses and the NF IRS array responses. Then, 
we formulate a sparse core tensor minimization problem with tri-modal log-sum sparsity constraints to tackle the NP-hard challenge.
Finally, the estimation method is further accelerated through higher-order singular value decomposition preprocessing, combined with the majorization-minimization strategy and a tailored tensor over-relaxation fast iterative shrinkage-thresholding technique. To evaluate the proposed algorithm, we derive the Cram\'{e}r-Rao lower bound and conduct the convergence analysis. Simulations demonstrate that the proposed scheme achieves an 13.6 dB improvement in terms of the normalized mean square error compared to benchmark methods with significantly reduced runtime.
\end{abstract}


\section{Introduction} \label{sec_1}
\IEEEPARstart{I}{ntelligent} reflecting surfaces (IRSs) are expected to be a driving force in overcoming propagation blockages and enhancing system spectral/energy efficiency in future sixth-generation (6G) communications~\cite{r1}. To enhance the potential benefits of IRS-assisted wireless systems, such as optimal beamforming design, the acquisition of accurate channel state information (CSI) is crucial~\cite{r2}. However, due to the lack of active transceivers for complex baseband processing, high-precision and low-overhead channel estimation for passive IRSs constitutes a major problem~\cite{r3}. This implies that separately estimating the user equipment (UE)-to-IRS channel and the IRS-to-base station (BS) channel is extremely difficult for fully passive IRSs~\cite{r4}. For this reason, estimating the cascaded channel, i.e., the equivalent UE-IRS-BS channel, was shown to be more convenient and equally effective in optimizing network performance~\cite{r5, r6}. Moreover, large-aperture IRSs are desirable in future wireless networks to reap the considerable array gain while alleviating the double-fading effect~\cite{r7}, which may increase the Rayleigh range of the IRS array. Thus, radio propagation under the extremely large-scale (XL)-IRS scenarios results in near-field (NF) radiation~\cite{r8}, where spherical wavefront-based channel model  relies not only on path angles but also on distances~\cite{r9}. 
Consequently, how to estimate the NF cascaded channel in a scalable manner is vital for the practical realization of XL-IRS systems.

\subsection{Related Works}
\textit{Far-Field (FF) Cascaded Channel Estimation}: Recently, many endeavors have been devoted to FF channel estimation techniques for IRS systems~\cite{r10, r11, r12, r13}.
In~\cite{r10}, partial active IRS elements were utilized for segmented channel estimation of both BS-IRS and IRS-UE links, where a compressed sensing (CS) solution was developed based on the millimeter-wave (mmWave) channel sparsity. However, active elements inevitably increases energy costs. The authors in ~\cite{r11} improved the conventional sparse Bayesian learning (SBL)-based cascaded channel estimation by imposing sparsity constraints on the hyperparameter update. Despite its accelerated convergence,  this method requires vectorization of the sparse cascaded channel, prohibiting scalability to XL-IRSs. 
The authors in~\cite{r12} proposed a hybrid channel estimation framework based on neural networks and CS, which improves estimation accuracy by jointly exploiting the common sparsity of subcarriers and the dual structural sparsity in the angle domain. Due to the uniform discretized grid-based dictionary design, angle quantization errors lead to severe grid mismatch effects. To avoid grid mismatch, a two-stage cascaded channel estimation scheme was proposed in~\cite{r13}. Initially, the column-block sparsity of the BS-IRS channel is leveraged for common subspace projection to estimate the angles of arrival (AoAs) at the BS. Then, the row-block sparsity of the IRS-UE channel enables a sparse recovery algorithm that avoids dictionary search and recovers the remaining channel parameters. {The authors of~\cite{a} proposed an IRS-enhanced coverage channel estimation algorithm based on power measurements. The core idea is to utilize channel spatial correlation to select typical user positions and model the received power using a single-layer neural network with network weights corresponding to the cascaded channel coefficients. This approach was further extended to wideband orthogonal frequency division multiplexing (OFDM) communication systems in ~\cite{bbb}, where the subcarrier average power is derived based on the channel's time-domain autocorrelation matrix, and the channel autocorrelation matrix is recovered through supervised learning using the power of reference signals from a subset of subcarriers.} {In \cite{Re2-16}, the method exploits the Vandermonde structure of the delay-frequency correlation factor matrix, which is specific to narrowband systems.}

\textit{NF Cascaded Channel Estimation for Uniform Linear Arrays (ULAs)}: Several studies have been carried out to explore the NF channel estimation methods for NF IRS systems~\cite{C-Z-21, r14, r15, r16, r17, r18}. {The method in \cite{C-Z-21} introduces independent anchor nodes to detect and calibrate the bidirectional visibility regions (VRs), then constrains the solution space to the identified VRs to reduce complexity, and recovers the cascaded channel through the ratio between the user-side cascaded channel and the anchor-node-side cascaded channel. However, adding independent anchor nodes incurs additional hardware overhead.} In~\cite{r14}, dictionary learning method was designed to adjust the column coherence of the NF channel codebook to enhance the NF sparse domain transformation. Then, a denoising convolutional neural network (DnCNN) was employed to tackle the grid leakage issue. 
To streamline the NF IRS cascaded channel estimation process, the authors in \cite{r15} proposed a two-stage approach. This approach first estimates the AoAs at the BS via atomic norm minimization, followed by Bayesian estimation of the IRS channel, which involves highly coupled parameters. This two-stage approach requires the design of specific time slot structures, which increases system complexity. Similarly, a two-stage CS framework was proposed in \cite{r16}, where the angles of departure (AoDs) estimation at the BS and the channel parameters estimation at the IRS  were converted into separate three-dimensional(3D) structured sparse recovery problems, which were solved using a 3D look ahead orthogonal matching pursuit (3D-LAOMP) algorithm. However, this approach requires a polar-domain codebook for sparse transformation of the NF channel, which increased higher computational complexity and codebook mismatch effect. In \cite{r17}, a redundancy-eliminating polar-domain codebook scheme was proposed to reduce the complexity of NF cascaded channel estimation. However, this method assumes configurations for ULA both the BS and IRS, and it cannot be directly extended to the more practical uniform planar array (UPA)-structured BS and IRS systems in 6G. %

\subsection{Motivations and Contributions}
Although recently a practical system with UPA-structured BS and IRS was studied in~\cite{r18}, it assumed a known BS-IRS channel and focused solely on estimating the IRS-UE channel. Cascaded channel estimation for the NF UPA-structured BS and IRS system faces the following challenges:
\textit{i) Codebook dimension explosion:} In the system with UPA-structured BS and IRS, the NF effects cause complex coupling among the elevation angle, azimuth angle, and distance, which leads to an overly expanded dictionary codebook space; \textit{ii) Quantization error accumulation:} The increased NF cascaded channel parameters in the UPA structure, combined with polar-domain codebook schemes, lead to accumulated quantization errors, degrading estimation performance.
Tensor-based methods have been shown to effectively address the complexity in NF XL-multiple-input multiple-output (MIMO) systems by separating channel components and enhancing estimation accuracy~\cite{ad0}.
Inspired by this, the authors in~\cite{t1} modeled the IRS cascaded channel estimation as a multilinear optimization problem based on Tucker tensor decomposition, where the BS-IRS and IRS-UE channels were represented as factor matrices and recovered via approximate message passing (AMP) method. This approach, however, neglects the inherent sparse structure of mmWave channels, which induces computational inefficiency in channel recovery.
{The tensor orthogonal matching pursuit
with joint search (T-OMP-JS)~\cite{ad1}  algorithm capitalizes on the spatial structural sparsity of the cascaded channel. However, its estimation process depends on discrete codebook matrices for each mode and employs OMP to identify non-zero entries in the core tensor. This reliance, however, becomes its primary bottleneck: the final estimation accuracy is fundamentally capped by the resolution of the codebooks, thereby preventing the algorithm from achieving high-precision results. The higher-dimensional rank-one approximation (HDR)~\cite{ad2} and two-stage higher-dimensional rank-one approximations (TSHDR)~\cite{ad3} methods lead to a fully coupled sparse structure across different modes by representing the channel tensor as a sum of multiple rank-one outer products. This coupling structure limits the model's ability to accurately capture the multi-dimensional sparsity of the channel and hinders the independent estimation of factor matrices, thus affecting the effective extraction of components in the cascade channel.}

In the practical system with UPA-structure BS and IRS, existing NF cascaded channel estimation schemes face the issues of codebook space explosion and cumulative quantization errors. To address the challenges, we propose an off-grid NF IRS cascaded channel estimation scheme based on tensor Tucker decomposition. The major contributions of this paper are summarized as follows:
\begin{itemize}
\item[$\bullet$] For the NF cascaded channel estimation, traditional channel matrix analysis-based codebook designs incur a three-order-of-magnitude increase in computational complexity for practical UPA structures. To overcome this, we propose a tensor modalization method that unifies 3D array responses through structured tensor construction. {Specifically, we exploit the tensor product of the horizontal/vertical response vectors of the UPA at the BS, and the IRS incident-reflective array response vector, to construct a 3D tensor.} Such 3D tensor's factor matrices correspond to multipath components of these array response vectors. 
This tensor-based method leverages the spatial structure of cascade channel, enabling efficient channel estimation through independent factor matrix estimation.

\item[$\bullet$] To avoid quantization error accumulation and excessive codebook size involved in on-grid estimation methods, we develop an off-grid Tucker decomposition-based cascade channel estimation framework. In this framework, the received signal is first modeled as a product of the sparse core tensor and three factor dictionary matrices. The non-zero entries of the sparse core tensor capture the path gain-delay components, while three factor matrices are spanned by the BS horizontal/vertical response vectors and the IRS incident-reflective array response vectors. To recover the cascaded channel, we formulate a sparse core tensor minimization problem with a reconstruction error constraint. A tri-modal log-sum term is introduced to replace the $\ell_0$ norm, addressing the NP-hard challenge.
 
\item[$\bullet$] To tackle the non-convexity of the constructed objective function and the involvement of many optimization variables, we employ the majorization-minimization (MM) algorithm to transform the original problem into a series of alternating iterative subproblems for estimating the sparse core tensor and factor matrices. To accelerate the algorithm convergence, we leverage higher-order singular value decomposition (HOSVD) to select the initial values for the sparse core tensor and factor matrices, and combine the tensor over-relaxed fast iterative shrinkage-thresholding (TOMFISTA) technique to further speed up the algorithm convergence. 

\item[$\bullet$] {We derive the Cram\'{e}r-Rao lower bound (CRLB) to assess the proposed algorithm. Furthermore, we reveal that when the IRS phase matrix adheres to the orthogonality constraint, the mean square error (MSE) of the proposed estimator converges to this lower bound. Numerical results validate that our approach surpasses benchmark methods in terms of normalized MSE (NMSE) and computational complexity.}
\end {itemize}

The remainder of the paper is structured as follows: Section \ref{sec_2} introduces the system and channel model for NF XL-IRS systems; Section \ref{sec_3} describes the off-grid Tucker decomposition-based cascade
channel estimation framework; Section \ref{sec_4} presents convergence analysis, CRLB analysis, and complexity analysis; Section \ref{sec_5}  provides simulation results and Section \ref{sec_6} concludes the paper.

\textit{Notations}: Bold-face upper- and lower- cases indicate matrices and vectors, respectively. ${\left(\cdot \right)^{\mathsf {T}}}$, ${\left(\cdot\right)^{\mathsf {H}}}$, $\left (\cdot \right )^*$, ${\left(\cdot  \right)^{-1}}$, and ${\left(\cdot\right)^\dag }$ stand for matrix transpose, conjugate transpose, conjugate, inverse and pseudoinverse, respectively. $\left\|\cdot\right\|_{0}$, $\left\|\cdot\right\|_{2}$ and ${\left\|\cdot\right\|_F}$ represent the $\ell_0$-norm, $\ell_2$-norm and Frobenius-norm, respectively. $\otimes$, $\odot$, $\bullet $, $\ast$ and $\circ$ denote the Kronecker product, Khatri-Rao product, row-wise Khatri-Rao product, Hadamard product and outer product, respectively. $\times_n$ denotes the $n$-mode product of the tensor and $\left \langle \cdot, \cdot \right \rangle$ denotes the inner product. $\operatorname{diag}(\cdot )$ represents diagonalization. ${\bf {I}}_{N}$ represents identity matrice of dimension ${N}$. ${\rm {vec}}\left ( \cdot \right )$ denotes vectorization. $\mathbb{E}\left [ \cdot  \right ] $ denotes the expectation operator. $\mathrm{Tr}\left ( \cdot  \right ) $ denotes the trace of a matrix.

\section{System Model} \label{sec_2}
\begin{figure}[t]
\centering
\includegraphics[width=2.8in]{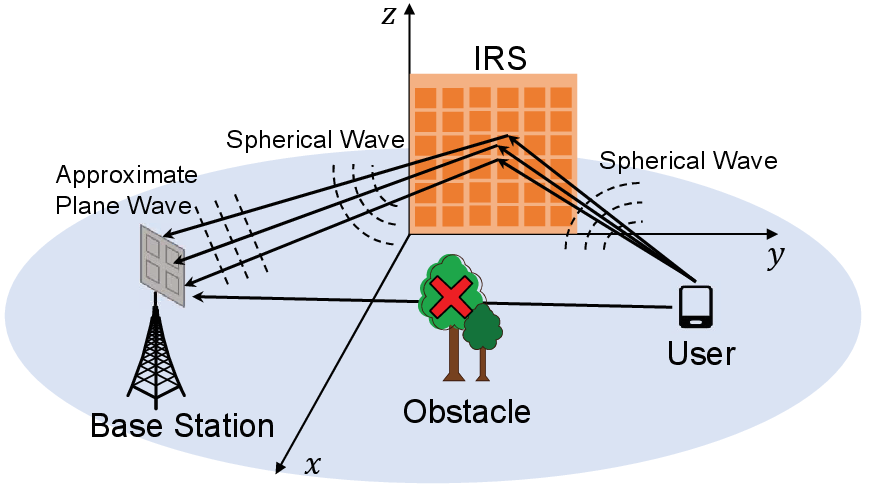}
\caption{NF XL-IRS-assisted communication systems.}
\label{fig_1}
\end{figure}

In this paper, we consider an uplink wideband orthogonal frequency-division multiplexing (OFDM) millimeter-wave system with single-antenna UEs, as shown in Fig.~\ref{fig_1}. Consider that the UE falls within the NF region of the IRS, while the BS is at a relatively far distance from the IRS. The BS is equipped with a UPA consisting of ${N_{\rm{b}}} = N_{\rm{z}} \times N_{\rm{y}}$ antennas, where $N_{\rm{z}}$ and $ N_{\rm{y}}$ denote the number of antennas in each row and column, respectively. 
The UPA-structured IRS comprises ${N_{\rm{r}}} = N_{\rm{r}}^{\rm{z}} \times N_{\rm{r}}^{\rm{y}}$ passive reflective elements, where $N_{\rm{r}}^{\rm{z}}$ and $N_{\rm{r}}^{\rm{y}}$ represent the number of elements in each row and column, respectively. 
The phase shift matrix for IRS elements is denoted by ${{\bm{\Omega }}} = {\rm diag}\left( \bf{v} \right)\in \mathbb{C}^{N_{\rm r}\times {N_{\rm r}}}$, where ${\bf v} = \left [ {{e^{j{\omega _1}}},{e^{j{\omega _2}}}, \ldots ,{e^{j{\omega _{{N_{\rm{r}}}}}}}}  \right ]^{\mathsf {T}}$, {and $\omega_j \in [0, 2\pi)$ denotes the reflection phase of the $j$-th IRS element.
    } Assume that the communication link between the BS and the UE is obstructed. ${{\bf{h}}_{{\rm{u}}}}$ and ${{\bf{H}}_{\rm{g}}}$ represent the UE-IRS and IRS-BS channels, respectively. { Additionally, both the BS and UE are located in the NF of the IRS.
}
 { In the IRS-BS channel, the reflected wave is radiated from the IRS and propagates toward the BS. For the IRS side, the reflected wave exhibits a noticeable spherical wavefront curvature across its large surface, and the propagation distance differences among array elements cannot be neglected. Therefore, a spherical-wave array response model is required to accurately characterize the spatial distribution of the reflected signal at the IRS side. In contrast, the BS array has a much smaller physical aperture compared with the extremely large-scale IRS surface, and the propagation distance differences among its antenna elements can be approximately ignored. As a result, the impinging wave on the BS can be well approximated as a plane wave,  as illustrated in Fig. 1, and thus a plane-wave array response model is adopted at the BS side. }

\subsection{Channel Model}
By appropriately deploying the BS and IRS, it is assumed that only a line-of-sight (LOS) link exists between the IRS and BS~\cite{r20}. 
For the XL-IRS channel, we adopt NF models as described in~\cite{r21}. 
Assume that $M$ subcarriers are utilized in the training protocol for channel estimation. The $m$-th subcarrier IRS-BS channel ${{\bf{H}}_{{\rm{g,}}m}}\in \mathbb{C}^{N_{\rm b} \times N_{\rm r}}$ is represented as 
\begin{equation}
{{\bf{H}}_{{\rm{g,}}m}} = {{\gamma }{e^{-j2\pi {f_{{m}}}{\xi }}}} {\bf{a}}_{{\rm{b}},m}\left( {{\psi_{{\rm{e}}}}},{{\varphi_{{\rm{a}}}}} \right){{\bf{a}}_{{\rm{r}},m}^{\mathsf{H}}}\left( {{\theta_{{\rm{e}}}},{\phi_{{\rm{a}}}},{u}} \right),
\label{eq_1}
\end{equation}
where ${{\gamma }}$ is the complex gain; ${{\psi_{{\rm{e}}}}}$ (or ${{\varphi_{{\rm{a}}}}}$) is the elevation (or azimuth)  angle of arrival (AoA) at the BS; ${{\theta_{{\rm{e}}}}}$ (or ${{\phi_{{\rm{a}}}}}$) is the elevation (or azimuth) angle of departure (AoD) at the IRS; ${\xi} = {u}/c$ is the time delay; $c$ is the light speed; and ${{u}}$ is the distance between the reference reflective element and the BS. Here, $f_m=f_c+\frac{2m-M}{2M}B$ is the baseband frequency of the $m$-th subcarrier, where $f_c$ is the carrier frequency and $B$ is the bandwidth.
Due to the wideband system, the array response is frequency-dependent~\cite{r22}. The spherical wavefront array response and the plane wavefront array response for the $m$-th subcarrier are denoted by ${{\bf{a}}_{{\rm{r}},m}}$ and ${\bf{a}}_{{\rm{b}},m}$, respectively.

The NF array response is affected by both the angle and distance. For the UPA configuration, the distance between the BS and the IRS needs to be calculated in a 3D structure~\cite{r23}. 
Accordingly, the wavelength difference from the BS to the $\left( {{n_{\rm{y}}},{n_{\rm{z}}}} \right)$-th IRS element is given by
\begin{align}
\Delta{u_{({n_{\rm{y}}},{n_{\rm{z}}})}} =& \left( {{{\left( {u\sin ({{\theta _{{\rm{e}}}}})\cos ({{\phi _{{\rm{a}}}}})} \right)}^2}}\notag\right. 
  \\ &+ {\left( {u\sin ({{\theta _{{\rm{e}}}}})\sin ({{\phi _{{\rm{a}}}}}) 
- \left( {{n_{\rm{y}}} - 1} \right)d} \right)^2}\notag
  \\&+ {\left. {{{\left( {u\cos ({{\theta _{{\rm{e}}}}}){n_{\rm{z}}}
 - \left( {{n_{\rm{z}}}
 - 1} \right)d} \right)}^2}} \right)^{1/2}}- u\notag\\
  = &\left( {{u^2} + {{\left( {{n_{\rm{y}}} - 1} \right)}^2}{d^2} + {{\left( {{n_{\rm{z}}} - 1} \right)}^2}{d^2}} \notag\right.
 \\&- 2\left( {{n_{\rm{y}}} - 1} \right)du\sin ({{\theta _{{\rm{e}}}}})\sin ({{\phi _{{\rm{a}}}}})
 \notag\\&{\ { - 2\left( {{n_{\rm{z}}} - 1} \right)du\cos ({{\theta _{{\rm{e}}}}})} \Big )^{1/2}}- u,
 \label{eq_2}
\end{align}
where $d$ is the antenna spacing. Based on the path difference~\cite{r24} form of the array response~\eqref{eq_2}, the NF array response ${{\bf{a}}_{{\rm{r}},m}}\left( {{\theta _{{\rm{e}}}},{\phi_{{\rm{a}}}},{u}} \right)\in \mathbb{C}^{N_{\rm r} \times 1}$ of the IRS at the $m$-th subcarrier is given by 
\begin{equation}
{{\bf{a}}_{{\rm{r}},m}}\left( {{\theta _{{\rm{e}}}},{\phi_{{\rm{a}}}},{u}} \right)= [{e^{ - j\frac{{2\pi {f_m}}}{c}\Delta {u_{(1,1)}}}},\cdots,{e^{ - j\frac{{2\pi {f_m}}}{c}\Delta {u_{{({{N_{\rm{r}}^{\rm {y}}},{N_{\rm{r}}^{\rm {z}}}})}}}}}]^{\mathsf {T}}.
\end{equation}

Since the antenna array at the BS  falls within the FF region of the IRS  \cite{r25}, the UPA's array response ${{\bf{a}}_{{\rm{b}},m}} \left( {{\psi _{{\rm{e}}}}},{{\varphi_{{\rm{a}}}}} \right)\in \mathbb{C}^{N_{\rm b} \times 1}$ at the $m$-th subcarrier is therefore written as
\begin{equation}
\begin{aligned}
{{\bf{a}}_{{\rm{b}},m}} \left( {{\psi _{{\rm{e}}}}},{{\varphi_{{\rm{a}}}}} \right) & 
=[1,\cdots,e^{-j2\pi(N_{\rm{y}}-1)\frac{f_m} {c}d \sin{{\psi_{{\rm{e}}}}}\sin{\varphi_{{\rm{a}}}}}]^{{\mathsf {T}}} \\
&\ \  \ \otimes[1,\cdots,e^{-j2\pi(N_{\rm{z}} -1)\frac{f_m} {c}d{\lambda}\cos\psi_{\mathrm{e}}}]^{{{\mathsf {T}}}} \\
&={\bf a}_{ N_{\rm{y}},m}(\psi_\mathrm{e},\varphi_\mathrm{a})\otimes{\bf a}_{N_{\rm{z}},m}(\psi_\mathrm{e}), 
\end{aligned}
\end{equation}
where $\lambda$ is the carrier wavelength; and ${\bf a}_{ N_{\rm{y}},m}(\psi_\mathrm{e},\varphi_\mathrm{a})\in \mathbb{C}^{N_{\rm y} \times 1}$ and 
${\bf a}_{N_{\rm{z}},m}(\psi_\mathrm{e})\in \mathbb{C}^{N_{\rm z} \times 1}$ denote the row and column array responses of the BS at the $m$-th subcarrier, respectively.

As the distance between the UE and the IRS is relatively short, the UE-IRS channel modeled as an NF form.  Hence, the UE-IRS channel at the $m$-th subcarrier is expressed as 
\begin{equation}
{{\bf{h}}_{{\rm{u,}}m}}  = \sum\limits_{l = 1}^L {{\alpha_l}{e^{ - j2\pi {f_{m}}{\zeta_l}}}} {{\bf{a}}_{{\rm{r}},m}}\left( {{\vartheta_{{\rm{e}},l}},{\varpi_{{\rm{a}},l}},{r_l}} \right)\in \mathbb{C}^{N_{\rm r} \times 1},
\label{eq_5}
\end{equation}
where $L$ represents the number of paths; ${\alpha_l}$ and ${\zeta_l}$ are the complex path gain and time delay, ${\vartheta_{{\rm{e}},l}}$ and ${\varpi_{{\rm{a}},l}}$ 
are the elevation and azimuth AoAs; and ${r_l}$ is the distance between the IRS reference element and the UE.
The expression of ${{\bf{a}}_{{\rm{r}},m}}$  defined in a similar manner.

Recalling~\eqref{eq_1} and~\eqref{eq_5}, the cascaded channel $\mathbf{G}_{m}\in \mathbb{C}^{N_{\rm b} \times {N_{\rm r}}}$ at the $m$-th subcarrier is reorganized as 
{\begin{align}
\mathbf{G}_{m}& =\mathbf{H}_{{\rm{g}},m}\operatorname{diag}({\bf h}_{{\rm{u}},m})\notag \\
&=\sum_{l=1}^{L}\beta_{l}e^{-j2\pi f_m \tau_{l}}{\bf a}{{_{{\rm {b}},m}}}\left( {{\psi _{{\rm{e}}}}},{{\varphi_{{\rm{a}}}}} \right) \notag\\&
\ \ \ \ \ \ \   \times ({{\bf{a}}^*_{{\rm{r}},m}}\left( {{\vartheta  _{{\rm{e}},l}},{\varpi  _{{\rm{a}},l}},{r_l}} \right) \ast   {{\bf{a}}_{{\rm{r}},m}}\left( {{\theta _{{\rm{e}}}},{\phi _{{\rm{a}}}},{u}} \right) )^{\mathsf {H}},
\label{eq_6}
\end{align}
where $\beta_l=\alpha_l\gamma$ represents the cascade channel gain of the $l$-th path, and the equivalent cascade delay of the $l$-th path is expressed as $\tau_{l}=\xi +\zeta_l$.}

\noindent{
\begin{remark}
The Rayleigh distance is typically used to quantify the boundary between the NF and FF regions. The Rayleigh distance $R$ is related to the effective aperture $D$ of the antenna array and the wavelength $\lambda$, and is given by $R = \frac{2D^2}{\lambda}$. In our extremely large IRS-assisted communication system model, the very large aperture of the IRS leads to a significantly increased Rayleigh distance. For example, in our simulation model, the number of IRS elements is ${N_{\rm{r}}} = N_{\rm{r}}^{\rm{z}} \times N_{\rm{r}}^{\rm{y}} =64\times4=256$, and the effective aperture is calculated as $D=\sqrt {((N_{\rm{r}}^{\rm{z}}-1)d)^2 ((N_{\rm{r}}^{\rm{y}}-1)d)^2}\approx 0.338 $ m. With a carrier frequency $f_c=28$ GHz,  the wavelength is $\lambda=c/f\approx 0.01071$ m. Substituting these values yields a Rayleigh distance of $R = \frac{2D^2}{\lambda}\approx 21.3$ m. The transmission distance between the BS and IRS is $7.2153$ m, and the transmission distance between the user and IRS follows ${\cal{U}}(5,10)$. These values are evidently much smaller than the Rayleigh distance, indicating that both links operate in the NF\footnote{{Due to the technological trends of large-aperture antennas and higher-frequency bands in 6G, NF propagation will become increasingly common~\cite{G1}, making NF modeling a practical requirement for fully unleashing the potential of 6G systems. NF modeling requires additional estimation of user positions, which can be addressed through hybrid positioning schemes (e.g., combining coarse global positioning system (GPS) measurements with advanced signal processing techniques) or by leveraging the integrated sensing and communication (ISAC) paradigm in 6G ~\cite{G2}.}} region of the IRS.
\end{remark}
} 

%\textcolor{blue}{ Due to the technological trends of large-aperture antennas and higher-frequency bands in 6G, NF propagation will become increasingly common~\cite{G1}, making near-field modeling a practical requirement for fully unleashing the potential of 6G systems. NF modeling requires additional estimation of user positions, which can be addressed through hybrid positioning schemes (e.g., combining coarse GPS measurements with advanced signal processing techniques) or by leveraging the integrated sensing and communication (ISAC) paradigm in 6G ~\cite{G2}.}

\subsection{Observation Signal Tensor Model}
\begin{figure*}[t]
\centering
\includegraphics[width=5.5in]{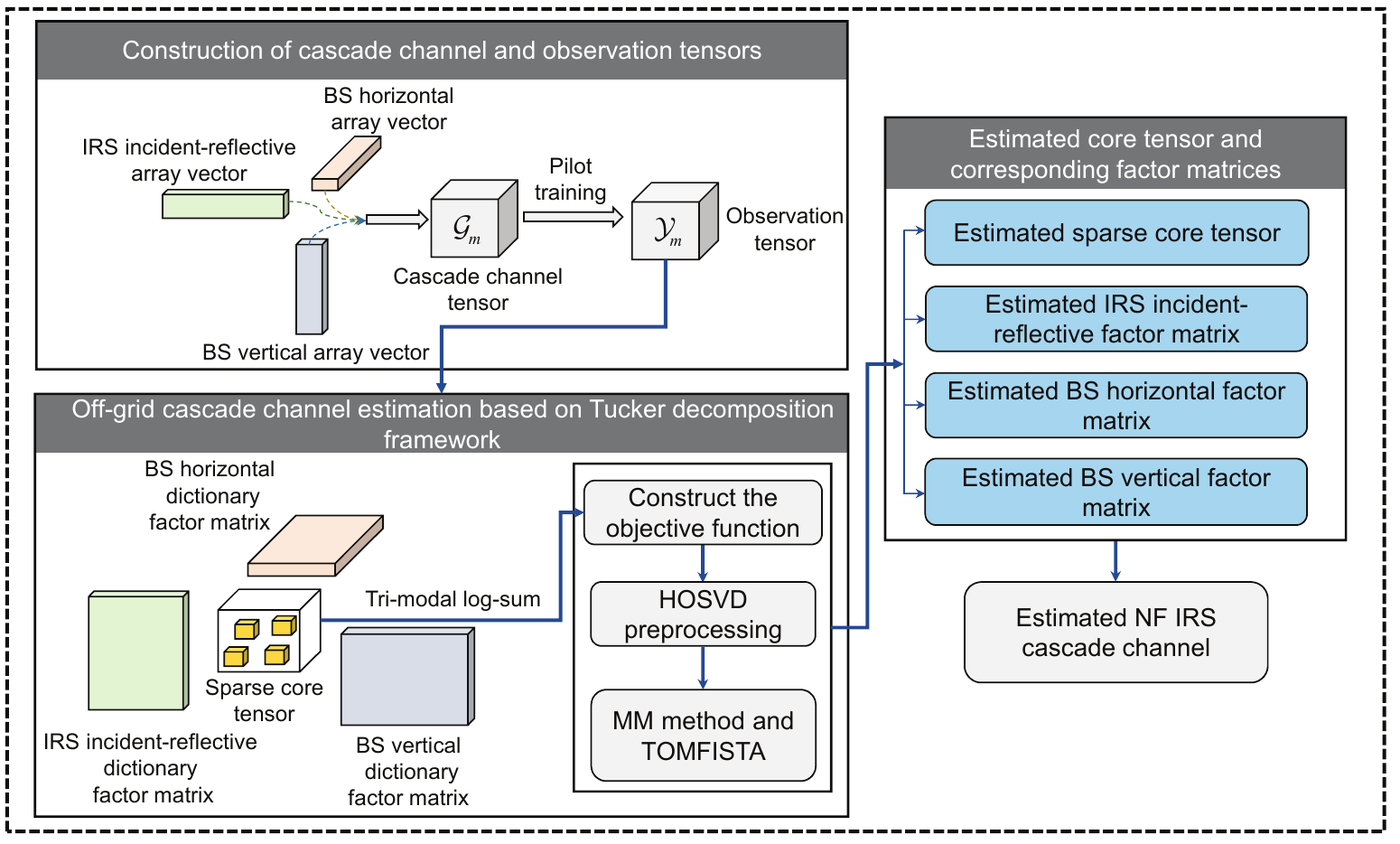}
\caption{Schematic diagram of the proposed NF IRS cascaded channel estimation algorithm based on Tucker decomposition: 1) Construct the cascaded channel and observation signal tensors; 2) Perform Tucker decomposition to formulate the objective function and apply an off-grid estimation approach to obtain the core tensor and factor matrices; 3) Based on the estimated core tensor and factor matrices, the estimated cascade channel is recovered.}
\label{fig_2}
\end{figure*}

We now present the developed tensor modalization method for NF XL-IRS channel estimation. The construction of the cascaded channel and observation signal tensor is illustrated in Fig.~\ref{fig_2}. { The scheme involves the following key steps: by taking the tensor product of the horizontal/vertical array response vectors of the UPA-structured BS with the IRS incident-reflective array response vector, a 3D tensor is formed to represent the NF IRS cascaded channel. Owing to the independence among these factor matrices, sparsity constraints can be imposed separately on each mode of the core tensor, enabling a more flexible and accurate characterization of the multi-dimensional sparse structure of the channel.} Subsequently, the observed signal is converted into tensor form by leveraging the properties of the tensor $n$-mode product. 

According to the tensor definition~\cite{r26}, the cascaded channel in Eqn.~\eqref{eq_6} can be restructured as a third-order tensor ${{{\bf{\cal G}}_m}}\in \mathbb{C}^{N_{\rm z} \times {N_{\rm r}}\times {N_{\rm y}}}$, as given by
\begin{equation}
{{{\bf{\cal G}}_m} = \mathop \sum \limits_{l = 1}^{{L}}\kappa_{l,m}  {\bf a}_{N_{\rm{z}},m}(\psi_\mathrm{e})\circ {{{{\bf b}}}_{m}}( {{{{\bf p}}_l}},{\bf q} )  \circ  {\bf a}_{ N_{\rm{y}},m}(\psi_\mathrm{e},\varphi_\mathrm{a})  },
\label{eq_7}
\end{equation}
where $\kappa_{l,m}=\beta_{l}e^{-j2\pi f_m \tau_{l}}$ denotes the cascaded gain-delay term at the $m$-th subcarrier along the $l$-th path, {${{{{\bf b}}}_{m}}\left ( {{\bf {p}}_l}, {{\bf {q}}}\right )={{\bf{a}}^*_{{\rm{r}},m}}\left( {{\vartheta  _{{\rm{e}},l}},{\varpi  _{{\rm{a}},l}},{r_l}} \right) \ast   {{\bf{a}}_{{\rm{r}},m}} \left( {{\theta _{{\rm{e}}}},{\phi _{{\rm{a}}}},{u}} \right)$ is the IRS incident-reflective array response for the $m$-th subcarrier} and the $l$-th path, ${{{\bf{p}}_l}}=\left [ {{\vartheta  _{{\rm{e}},l}},{\varpi  _{{\rm{a}},l}},   {r_l}} \right ]$ collects the IRS incident array response parameters for the $l$-th path, and ${\bf{q}}=\left [{{\theta _{{\rm{e}}}},{\phi _{{\rm{a}}}},{u}} \right ]$ collects the IRS reflection array response parameters.

In the uplink channel estimation phase, $P$ consecutive OFDM symbols are utilized for training. Let ${s_{m,p}}$ represent the pilot of the $p$-th OFDM symbol on the $m$-th subcarrier. 
Then, the received signal ${{\bf y}_{m,p}}$ of the $p$-th OFDM symbol on the $m$-th subcarrier is expressed as
\begin{equation}
{{\bf y}_{m,p}} = {{\bf{G}}_{m}}{{\bf{v}}_p}s_{m,p} + {{\bf n}_{m,p}}\in \mathbb{C}^{N_{\rm b}\times {1}},
\label{eq_8}
\end{equation}
where ${{\bf{v}}_p}\in \mathbb{C}^{N_{\rm r}\times {1}}$ is the IRS reflection vector for the $p$-th OFDM symbol, ${{\bf n}_{m,p}}\in \mathbb{C}^{N_{\rm b}\times {1}}$ is the additive Gaussian white noise with its elements  following ${\cal C} {\cal N}\sim  (0,\sigma ^2)$, and $\sigma^2$ is the noise power.

To simplify the calculation, we set $s_{m,p}=1$ for all indexed values. For the $m$-th subcarrier, the received signals  of $P$ collected OFDM symbols are obtained as
\begin{equation}
\mathbf{Y}_m=[\mathbf{y}_{m,1},\ldots,\mathbf{y}_{m,P}]=\mathbf{G}_m\mathbf{V}+\mathbf{N}_m,
\label{eq_9}
\end{equation}
where $\mathbf{Y}_m\in \mathbb{C}^{N_{\rm b}\times {P}}$, $\mathbf{V}=[{\bf v}_{1},\ldots,{\bf v}_{P}]\in \mathbb{C}^{N_{\rm r}\times {P}}$ and ${\mathbf N}_m=[{\bf n}_{m,1},\ldots,{\bf v}_{m,P}]\in \mathbb{C}^{N_{\rm b}\times {P}}$.

According to the properties of the tensor $n$-mode product \cite{Tda}, and recalling Eqns.~\eqref{eq_6} and~\eqref{eq_7}, the observation signal $\mathbf{Y}_m$ in~\eqref{eq_9} transformed into a tensor form as follows.
\begin{equation}
{{{\bf{\cal Y}}_m} = {{\bf{\cal G}}_m}\times_2 {\bf{V}}^{\mathsf {T}}}+{\cal N}_m \in \mathbb{C}^{N_{\rm z} \times P\times {N_{\rm y}}},
\label{eq_10}
\end{equation}
where ${{\cal N}_m}\in \mathbb{C}^{N_{\rm z} \times P\times {N_{\rm y}}}$ denotes the noise tensor.

\section{Off-Grid Tucker Decomposition-Based NF Cascaded Channel Estimation Framework} \label{sec_3}
To avoid computationally expensive tensor processing, we introduce a Tucker decomposition method to enable dimensionality reduction while extracting latent low-rank structures and denoising. Considering the inherent sparsity of mmWave channels, this section presents an off-grid NF cascaded channel estimation framework based on Tucker decomposition, as shown in Fig.~\ref{fig_2}. 
The framework models the received signal as the mode product of a sparse core tensor and three factor dictionary matrices, where the nonzero entries of the sparse core tensor represent the path gain-delay components, and the factor dictionary matrices are spanned by the BS horizontal/vertical and IRS incident-reflection array response vectors, respectively. To recover the cascaded channel, we propose a sparse core tensor minimization problem with a reconstruction error constraint and introduce a tri-modal log-sum term to replace the $\ell_0$ norm, thus solving the NP-hard problem. To address the non-convexity of the formulated optimization problem, we devise an MM algorithm to transform the original problem into four alternating iterative subproblems. Finally, we utilize the HOSVD to select initial values and combine the TOMFISTA technique to accelerate the algorithm's convergence. The dimensionality reduction of the factor matrices through HOSVD decomposition allows the off-grid scheme to optimize within a very small continuous factor matrix space, significantly reducing the number of parameters that need to be computed by combining the sparse structure prior of the core tensor. In contrast, the on-grid method requires traversing a large dictionary codebook of discrete points, which increases the computational cost.

{In fact, our proposed scheme achieves off-grid processing. {The factor dictionary matrices mentioned in \eqref{eq_r12}--\eqref{eq_14} serve only as initial codebooks.} Our scheme does not select the true solution from these initial codebooks; instead, it models the channel sparsity within the core tensor of a Tucker structure, and iteratively optimizes both the factor dictionary matrices and the core tensor to realize continuous estimation of the cascaded channel. In contrast, the atomic norm minimization (ANM) method defines the signal model over a continuous atomic set and seeks the minimum atomic norm solution within an infinite-dimensional convex set through semidefinite programming (SDP)~\cite{Ra}. Although such methods perform continuous off-grid estimation, they suffer from the curse of dimensionality in high-dimensional scenarios (e.g., extremely large-scale IRS systems), as both the construction of the atomic set and the complexity of SDP optimization become prohibitively high. Another category, the off-grid sparse Bayesian learning (off-SBL) methods, typically perform Bayesian inference directly on the continuous-domain channel vector or introduce parameter offsets to correct the discretized dictionary model~\cite{Rs2, Rs1}. The method in ~\cite{Rs2} requires full vectorization of the observed signals and Bayesian inference under a high-dimensional sparse dictionary, leading to enormous computational and memory costs. The method in ~\cite{Rs1} introduces continuous offset corrections on a discrete grid. However, it faces challenges in NF IRS scenarios, where the incident-reflected angles and propagation distances are strongly and nonlinearly coupled. This coupling makes the modeling of offset parameters difficult.}

Initially, we exploit the sparse nature of the mmWave channel to convert the cascaded channel into a sparse form. According to \eqref{eq_7}, the cascaded channel is expressed as a summation of combinations of three rank-one tensors corresponding to multiple propagation paths. Given that mmWave channels typically contain only a limited number of effective propagation paths, we construct three  factor dictionary matrices based on these rank-one components, and concentrate the path energy at a few specific indices within the core tensor, thereby forming a sparse core tensor structure. The sparse transformation of \eqref{eq_7} can be written as
\begin{equation}
{{{\bf{\cal G}}_m} = {\cal {Z} }_{m}\times_1 {{\bf{A}}}_{{\rm{z}},m }\times_2 {\bf{B}}_{{\rm{r}},m}\times_3{\bf{A}}_{{\rm{y}},m}},
\label{eq_11}
\end{equation}
where ${{{\bf{\cal Z}}_m}}\in \mathbb{C}^{G_{\rm z} \times {G_{\rm r}}\times {G_{\rm y}}}$ represents the sparse core tensor at the $m$-th subcarrier. The
 non-zero elements of ${{{\bf{\cal Z}}_m}}$ are denoted by $\kappa_{l,m}$,  $1\le l\le L,1\le m\le M$. {${\bf{A}}_{{\rm{z}},m }\in \mathbb{C}^{N_{\rm z} \times {G_{\rm z}}}$, ${\bf{B}}_{{\rm{r}},m}\in \mathbb{C}^{N_{\rm r} \times {G_{\rm r}}}$, and ${\bf{A}}_{{\rm{y}},m }\in \mathbb{C}^{N_{\rm y} \times {G_{\rm y}}}$ are the 
factor dictionary matrices,} i.e.,
\begin{align}
{\bf{A}}_{{\rm{z}},m } &= \left[ {{\bf a}_{N_{\rm{z}},m}({\Bar\psi}_\mathrm{e,1}), \ldots ,{{\bf a}_{N_{\rm{z}},m}({\bar \psi}_{\mathrm{e},G_{{\rm z}}})}} \right],
\label{eq_r12}
\end{align}
\begin{align}
{\bf{B}}_{{\rm{r}},m} &= \left[ {{{{\bf{b}}}_{m}}( {{{\bar{\bf{ p}}}_1}},{\bar {\bf{ q}}}_1 ), \ldots ,{{{\bf{b}}_m}}( {{\bar{{\bf{ p}}}_{G_{\rm r}}}},{\bar {\bf{ q}}}_{G_{\rm{r}}} )} \right],
\label{eq_r13}
\end{align}
\begin{align}
{\bf{A}}_{{\rm{y}},m}& = \left[ {{\bf a}_{ N_{\rm{y}},m}({\bar \psi}_\mathrm{e,1},{\bar \varphi}_\mathrm{a,1}), \ldots ,{\bf a}_{ N_{\rm{y}},m}({\bar \psi}_{\mathrm{e},G_{{\rm y}}},{\bar{\varphi}}_{\mathrm{a},G_{{\rm y}}})} \right].
\label{eq_14}
\end{align}
where $\left \{ {\bar \psi}_\mathrm{e},{\bar \varphi}_\mathrm{a} ,{{\bar{\bf{ p}}}},{\bar {\bf{ q}}} \right \} $  collects the parameters of the angle and distance codebooks~\cite{r27,r28}. $G_{{\rm z}} $, $G_{{\rm r}}$ and $G_{{\rm y}}$  representing the sizes of the factor matrix dictionaries, respectively. {It is worth emphasizing that the matrices ${\bf{A}}_{{\rm{z}},m }$, ${\bf{B}}_{{\rm{r}},m}$ and ${\bf{A}}_{{\rm{y}},m}$ are dense factor dictionary matrices that collect all possible array response patterns and do not exhibit sparsity themselves. Accordingly, ${\bf{A}}_{{\rm{z}},m }$, ${\bf{B}}_{{\rm{r}},m}$ and ${\bf{A}}_{{\rm{y}},m}$ serve as the bases for sparse representation instead of sparse matrices.} {The nonzero entries of ${{{\bf{\cal Z}}_m}}$ correspond solely to the cascaded gain-delay terms of effective paths, while all other entries are zero, yielding a sparse core tensor ${{{\bf{\cal Z}}_m}}$. The locations of the sparse nonzero entries in ${\bf{\cal Z}}_m$  generally correspond to the indices of the actual array responses within the three factor dictionary matrices. These array response vectors correspond to the directions of the effective propagation paths in the physical channel. Consequently, the sparse representation in ${\bf{\cal Z}}_m$ explicitly captures the inherent physical sparsity of the cascaded channel, while the combination of ${\bf{\cal Z}}_m$ and the associated dictionary entries fully preserves the complete information of the   ${\bf{\cal G}}_m$.}

Substituting \eqref{eq_11} into \eqref{eq_10}, the observed tensor signal with a sparse structure is rewritten as
\begin{equation}
{{{\bf{\cal Y}}_m} ={{\cal Z}_{m}\times_1 {\bf{A}}_{{\rm{z}},m }\times_2 ({\bf{V}}^{\mathsf {T}}{\bf{B}}_{{\rm{r}},m})\times_3{\bf{A}}_{{\rm{y}},m}} }+{\cal N}_m.
\label{eq_15}
\end{equation}

{The observation signal model in \eqref{eq_15} adopts a Tucker decomposition-based approach, mainly because this model can better characterize the multi-dimensional structured sparsity of channels in XL-IRS communication systems, thereby achieving more stable, interpretable, and accurate channel reconstruction.}

\subsection{Problem Formulation}
Through the above sparse representation of the signal tensor, we can further transform the NF cascaded channel estimation problem into a \textit{sparse tensor recovery problem}.
For ease of illustration, we take the $m$-th subcarrier as an example and subsequent optimization problems are formulated considering only the $m$-th subcarrier. 
From Eqn.~\eqref{eq_11}, the sparse core tensor ${\cal Z}_{m}$ is shown to possess a 3D sparse structure. Then, by imposing sparsity constraints to each dimension of ${\cal Z}_{m}$, the tensor sparse recovery problem can be formulated as
\begin{subequations}
\label{eq_16}
\begin{align}
 & \qquad \qquad   \mathop  {\min }\limits_{{{\cal Z}_m}, \left \{ {{\bf{A}}^{(n)}_{m}} \right \} } \;  \sum\limits_{n = 1}^{N_{\rm {d}}} \parallel  {{\bf{z}}_{m,n}}{\parallel _0}  
  \label{eq_16a}\\
&  {\rm{s}}.{\rm{t}}.\left\| {{{\cal Y}_m} - {{\cal Z}_{m}\times_1 {{\bf{A}}^{(1)}_{m}}\times_2 {{\bf{A}}^{(2)}_{m}}\times_3{{\bf{A}}^{(3)}_{m}}}} \right\|_F^2 \le \varepsilon ,
\label{eq_16b}
\end{align}
\end{subequations}
where ${N_{\rm {d}}}=3$ denotes the tensor dimension. For notational convenience, we define: 
${{\bf{A}}^{(1)}_{m}}={\bf{A}}_{{\rm{z}},m}, {{\bf{A}}^{(2)}_{m}}={{\bf{V}}^{\mathsf {T}}{\bf{B}}_{{\rm{r}},m} }, {{\bf{A}}^{(3)}_{m}}={\bf{A}}_{{\rm{y}},m }$. $\varepsilon $ is a tolerance parameter related to noise statistics. 
${{\bf{z}}_{m,n}}$ represents the vector comprising the $\ell_2$-norms of the row vectors obtained from the $n$-mode unfolding of ${\cal Z}_{m}$, with its $i$-th entry expressed as
\begin{equation}
z_{m,n,i}\triangleq\parallel\mathcal{Z}_{m,n,i}\parallel_2.
\label{eq_17}
\end{equation}

The optimization problem \eqref{eq_16}, which involves the $\ell _0$-norm, is NP-hard~\cite{r29}. To address this, we approximate the 3D sparsity constraints of the core tensor. Given that the log-sum function serves as a relaxation of the $\ell_0$-norm~\cite{r30}, we employ it to approximate the sparse core tensor across each dimension, thereby forming a tri-modal log-sum constraint. Consequently, \eqref{eq_16} is reformulated as
\begin{subequations}
\label{eq_18}
\begin{align}
& \mathop {\min }\limits_{{{\cal Z}_m}, \left \{ {{\bf{A}}^{(n)}_{m}} \right \}}  \; L_{\rm z}\left({\cal{Z}}_m\right)= \sum\limits_{n = 1}^{N_{\rm {d}}} \sum\limits_{i = 1}^{I_{n}} {\rm {log}}( \parallel  {{\cal{Z}}_{m,n,i}}{\parallel ^{2} _2} +\delta )\label{eq_18a}\\
& \quad {\rm{s}}.{\rm{t}}.\left\| {{{\cal Y}_m} - {{\cal Z}_{m}\times_1 {{\bf{A}}^{(1)}_{m}}\times_2 {{\bf{A}}^{(2)}_{m}}\times_3{{\bf{A}}^{(3)}_{m}}}} \right\|_F^2 \le \varepsilon,
\label{eq_18b}
\end{align}
\end{subequations}
where $I_{n}$ is the number of rows in the matrix obtained from $n$-mode unfolding of the tensor, ${\rm {log}}(\cdot)$ is the logarithmic function, and {$\delta$ is introduced as a small positive constant to prevent ill-conditioned numerical behavior in the logarithmic operation under extreme cases, such as when the argument of the log-sum sparsity surrogate approaches zero and the logarithmic term tends to negative infinity. }

%and $\delta$ is a small positive parameter to prevent numerical instability in logarithmic operations.

To dissolve problem \eqref{eq_18}, we adopt the Tikhonov regularization method to transform the constrained optimization problem into an unconstrained optimization problem, as given by
\begin{align}
& \mathop{\min }\limits_{{{\cal Z}_m}, \left \{ {{\bf{A}}^{(n)}_{m}} \right \}}  \; L_{\rm {o}}\left (  {{{\cal Z}_m}, \left \{ {{\bf{A}}^{(n)}_{m}} \right \}} \right ) = \sum\limits_{n = 1}^{N_{\rm {d}}} \sum\limits_{i = 1}^{I_{n}} {\rm {log}}( \parallel  {{\cal{Z}}_{m,n,i}}{\parallel ^{2}_2} +\delta )\notag \\
& \qquad \qquad\qquad+\lambda_1  \left\| {{{\cal Y}_m} - {{\cal Z}_{m}\times_1 {{\bf{A}}^{(1)}_{m}}\times_2 {{\bf{A}}^{(2)}_{m}}\times_3{{\bf{A}}^{(3)}_{m}}}} \right\|_F^2 \notag \\
&\qquad\qquad\qquad\qquad+\lambda_2\sum_{n=1}^{N_{d}} \left \|  {{\bf{A}}^{(n)}_{m}} \right \|^2_F,
\label{eq_19}
\end{align}
where $L_{\rm {o}}\left({{{\cal Z}_m}, \left \{ {{\bf{A}}^{(n)}_{m}} \right \}} \right )$ represents the loss function, and $\lambda_1$ and $\lambda_2$ are regularization parameters. The first regularization term (the second item on the right-hand side) penalizes the fitting error. By tuning $\lambda_1$, the model's data fitting error can be controlled. The third item on the right-hand side acts as the second regularization term (associated with $\lambda_2$) aims to prevent the factor matrices from becoming excessively large or overly complex, thereby enhancing the model's generalization ability.

\subsection{Cascaded Channel Recovery Algorithm}
Given that the objective function in \eqref{eq_19} is non-convex and involves a large number of optimization variables, we employ the MM algorithm to reformulate the problem into a sequence of tractable convex subproblems. This transformation from non-convex to convex is accomplished by replacing a surrogate function.

Since the log-sum term in the original objective function \eqref{eq_19} is nonconvex, a convex upper bound function for the log-sum term can be constructed based on first-order Taylor expansion, i.e., 
\begin{align}
 &{\rm {log}}( \parallel {{\cal{Z}}_{m,n,i}}{\parallel^{2} _2} +\delta )\le \frac{\parallel  {{\cal{Z}}_{m,n,i}}{\parallel^{2} _2} +\delta}{\parallel  {{{\cal{Z}}^{[t]}}_{m,n,i}}{\parallel^{2} _2} +\delta}\notag 
\\ &  \qquad \qquad \qquad \qquad \quad  +{\rm {log}}( \parallel  {{\cal{Z}}^{[t]}_{m,n,i}}{\parallel^{2} _2} +\delta )-1,
\label{eq_20}
\end{align}
where the superscript $[t]$ is the number of iterations. The equality in \eqref{eq_20} holds when ${{{\cal{Z}}}_{m}} = {{\cal{Z}}_m^{[t]}}$.
In \eqref{eq_19}, the log-sum term is nested within a double summation. According to \eqref{eq_20}, we can derive the double-sum form $f\left ( {{{\cal{Z}}}_{m}}  \mid {{{\cal{Z}}^{[t]}_m}} \right )$ for the surrogate function, as given by
\begin{align}
&f\left ( {{{\cal{Z}}}_{m}}  \mid {{{\cal{Z}}^{[t]}_m}} \right )= \langle{{{\cal{Z}}}_{m}},\mathcal{D}_{m}^{[t]}\ast {{{\cal{Z}}}_{m}}\rangle \notag \\
& \qquad\qquad+\sum_{n=1}^{N_{\rm d}}\sum_{i=1}^{I_n}{\rm {log}}( \parallel  {{\cal{Z}}^{[t]}_{m,n,i}}{\parallel^{2} _2} +\delta )-\sum_{n=1}^{N_d}I_n,
\label{eq_21}
\end{align}
where $\mathcal{D}_{m}^{[t]}\in \mathbb{C}^{G_{\rm z} \times {G_{\rm r}}\times {G_{\rm y}}}$. The $\left ( i_1,i_2,i_3 \right )$-th element of $\mathcal{D}_{m}^{[t]}$ can be expressed as
\begin{align}
\mathcal{D}_{m,i_1i_2i_3}^{[t]}=\sum_{n=1}^{N_{\rm d}}\left(\left\|\mathcal{Z}_{(m,n,i_n)}^{[t]}\right\|_F^2+\delta \right)^{-1}.
\label{eq_22}
\end{align}
where $1\le i_1\le G_{\rm z}$, $1\le i_2\le G_{\rm r}$, $1\le i_3\le G_{\rm y}$.

Substituting the surrogate of \eqref{eq_21} into \eqref{eq_19}, the new objective is reformulated as
\begin{align}
 & S_{\rm {n}}\left (  {{{\cal Z}_m}, \left \{ {{\bf{A}}^{(n)}_{m}} \right \}}\mid  {{{\cal{Z}}^{[t]}_m}} \right ) = f\left ( {{{\cal{Z}}}_{m}}  \mid {{{\cal{Z}}^{[t]}_m}} \right )\notag \\
& \qquad \qquad\qquad+\lambda_1  \left\| {{{\cal Y}_m} - {{\cal Z}_{m}\times_1 {{\bf{A}}^{(1)}_{m}}\times_2 {{\bf{A}}^{(2)}_{m}}\times_3{{\bf{A}}^{(3)}_{m}}}} \right\|_F^2 \notag \\
&\qquad\qquad\qquad\qquad+\lambda_2\sum_{n=1}^{N_{{\rm d}}} \left \|  {{\bf{A}}^{(n)}_{m}} \right \|^2_F.
\label{eq_23}
\end{align}

Following the MM principle, solving problem \eqref{eq_19} can be recast into iteratively minimizing \eqref{eq_23}. 
Since the optimization variables in \eqref{eq_23} include the core tensor and the three factor matrices,
we adopt an alternating minimization strategy to solve for each variable.

We first solve for the variable ${{\cal Z}_m}$, with the factor matrix $\left \{{{\bf{A}}^{(n)}_{m}}\right \}$ fixed. The optimization of ${{\cal Z}_m}$ with respect to \eqref{eq_23} can be reduced into 
\begin{align}
 & \mathop  {\min }\limits_{{{\cal Z}_m} } \quad {\cal {F}}({\cal Z}_m)= \langle{{{\cal{Z}}}_{m}},\mathcal{D}_{m}^{[t]}\ast{{{\cal{Z}}}_{m}}\rangle\notag \\
& +\lambda_1  \left\| {{{\cal Y}_m} - {{\cal Z}_{m}\times_1 {{\bf{A}}^{(1)}_{m}}\times_2 {{\bf{A}}^{(2)}_{m}}\times_3{{\bf{A}}^{(3)}_{m}}}} \right\|_F^2 .
\label{eq_24}
\end{align}

Conventional methods for solving \eqref{eq_24} require vectorizing each variable and applying least squares. However, this vectorization significantly increases dimensionality, leading to substantial computational inefficiency and prohibitive delays~\cite{r33}. To address this, we develop an accelerated algorithm, named TOMFISTA, for estimating the sparse core tensor ${{{\cal Z}_m} }$. In particular, we first rewrite the objective in \eqref{eq_24} as
\begin{align}
&     \mathop  {\min }\limits_{{{\cal Z}_m} } \quad {\cal {F}}({\cal Z}_m)={\cal {F}}_{1}({\cal Z}_m)+{\cal {F}}_{2}({\cal Z}_m),
\label{eq_25}
\end{align}
where 
\begin{align}
{\cal {F}}_{1}({\cal Z}_m)=\langle{{{\cal{Z}}}_{m}},\mathcal{D}_{m}^{[t]}\ast {{{\cal{Z}}}_{m}}\rangle,
\label{eq_26}
\end{align}
and
\begin{align}
{\cal {F}}_{2}({\cal Z}_m)=\lambda_1  \left\| {{{\cal Y}_m} - {{\cal Z}_{m}\times_1 {{\bf{A}}^{(1)}_{m}}\times_2 {{\bf{A}}^{(2)}_{m}}\times_3{{\bf{A}}^{(3)}_{m}}}} \right\|_F^2. 
\label{eq_27}
\end{align}

Due to its simplicity and applicability in large-scale problems,
the TOMFISTA is leveraged to solve \eqref{eq_25} through the following iterative sequence
\begin{align}
{{\cal K}_{m}^{[k]}} = {\rm{prox}}{_{\lambda_3 {\cal {F}}_{1}}}({{\cal W}_{m}^{[k]}} - \lambda_3  \nabla {{\cal F}_{2}}\left( {{{\cal W}_{m}^{[k]}}} \right)),
\label{eq_28}
\end{align}
where ${\rm{prox}}{_{\lambda_3 {\cal {F}}_{1}}}(\cdot )$ represents the proximal operator \cite{r34}; the intermediate variable ${{\cal W}_{m}^{[k]}}$ is initialized with ${{\cal W}_{m}^{[1]}}={{\cal Z}_{m}^{[0]}}$; ${{\cal Z}_{m}^{[0]}}$ denotes the initial value of ${{\cal Z}_{m}^{}}$; $\lambda_3$ is the step size of the gradient descent; and $\nabla {{\cal F}_{2}}\left( {{{\cal W}_{m}^{[k]}}} \right))$ represents the gradient of $ {{\cal F}_{2}}\left( {{{\cal W}_{m}^{[k]}}} \right))$, i.e.,
\begin{align}
\nabla {{\cal F}_{2}}\left( {{{\cal W}_{m}^{[k]}}} \right) = &-2 \lambda_1 \left( \mathcal{Y}_m - {\cal Z}_m^{[k]} \times_1 \mathbf{A}_m^{(1)} \times_2 \mathbf{A}_m^{(2)} \times_3 \mathbf{A}_m^{(3)} \right) \notag \\
&\times_1 (\mathbf{A}_m^{(1)})^{\mathsf {H}} \times_2 (\mathbf{A}_m^{(2)})^{\mathsf {H}} \times_3 (\mathbf{A}_m^{(3)})^{\mathsf {H}}.
\label{eq_29}
\end{align}

According to the proximal operator method, the solution to \eqref{eq_28} can be derived as
\begin{align}
({{\cal K}_{m}^{[k]}})_{{i_1}{i_2}{i_3}} =\frac{({{\cal W}_{m}^{[k]}} - \lambda_3  \nabla {{\cal F}_{2}}( {{{\cal W}_{m}^{[k]}}}))_{{i_1}{i_2}{i_3}}}{(2\lambda_3\mathcal{D}_{m}^{[t]}+{\cal I})_{{i_1}{i_2}{i_3}}}, 
\label{eq_30}
\end{align}
where $\left ( \cdot  \right )_{{i_1}{i_2}{i_3}}$ denotes the $({{i_1},{i_2},{i_3}})$-th entry of the tensor, and ${\cal I}$ denotes a tensor with the same dimensions as $\mathcal{D}_{m}^{[t]}$, with all elements set to one. To guarantee a monotonic decrease in the objective function throughout the iterations, the following manipulation is applied.
\begin{align}
{{\cal Z}_{m}^{[k]}} = \arg \min \left\{ {{\cal F}({\cal J}_{m})|{\cal J}_{m} \in \left\{ {{{\cal K}_{m}^{[k]}},{{\cal Z}_{m}^{[k - 1]}}} \right\}} \right\}.
\label{eq_31}
\end{align}

As a result, the intermediate variable ${{{\cal W}_{m}^{[k]}}}$ in \eqref{eq_28} can be updated by
\begin{align}
{{\cal W}_{m}^{[k + 1]}} \buildrel \over = {{\cal Z}_{m}^{[k]}}& + \frac{{{\eta ^{[k]}}}}{{{\eta ^{[k + 1]}}}}({{\cal K}_{m}^{[k]}} - {{\cal Z}_{m}^{[k]}})\notag \\
  & + \frac{{{\eta ^{[k]}} - 1}}{{{\eta ^{[k + 1]}}}}({{\cal Z}_{m}^{[k]}} - {{\cal Z}_{m}^{[k - 1]}})\notag \\
 &+ \frac{{{\eta ^{[k]}}}}{{{\eta ^{[k + 1]}}}}(1-\rho )({{\cal W}_{m}^{[k]}} - {{\cal K}_{m}^{[k]}}),
 \label{eq_33}
\end{align}
where ${\eta ^{[k + 1]}} = \frac{{1 + \sqrt {1 + 4{{\left( {{\eta ^{[k]}}} \right)}^2}} }}{2}$ is an acceleration factor~\cite{r33}, and $\rho$ is an iteration-dependent balancing parameter. {The monotonic selection procedure in (31) (i.e., selecting at each iteration the core-tensor solution that minimizes the objective function) will not cause the iterations to stagnate or become non-convergent. Its synergy with the acceleration step of the algorithm, i.e., Eq. (32), can prevent oscillations or divergence induced by excessive extrapolation, thereby ensuring that the quality of the core tensor solution keeps improving with the iterations.}

Following the sequence of \eqref{eq_30}–\eqref{eq_33} through ${k_{\text {max}}}$ iterations, we have ${{\cal Z}_{m}^{}}={{\cal Z}_{m}^{[k_{\text {max}}]}}$. Given ${{\cal Z}_{m}^{}}$, we optimize remaining three factor matrices, i.e., ${{\bf{A}}^{(1)}_{m}}$, ${{\bf{A}}^{(2)}_{m}}$ and ${{\bf{A}}^{(3)}_{m}}$ based on the alternating minimization strategy. Recalling \eqref{eq_23}, the objective function for solving each factor matrix is given by
\begin{align}
\mathop{\min }\limits_{{\left \{{{\bf{A}}^{(n)}_{m}}\right \}} } \quad  
&\lambda_1  \left\| {{{\cal Y}_m} - {{\cal Z}_{m} \prod_{n\ne k} \times_n {{\bf{A}}^{(n)}_{m}}}} \right\|_F^2 
+\lambda_2\sum_{n=1}^{N_{\rm d}} \left \|  {{\bf{A}}^{(n)}_{m}} \right \|^2_F ,
\label{eq_34}
\end{align}
where 
\begin{align}
&{\cal Z}_{m} \prod_{n\ne k} \times_n {{\bf{A}}^{(n)}_{m}}=\mathcal{X}\times_1{{\bf{A}}^{(1)}_{m}}\cdots\times_{k-1}{{\bf{A}}^{(k-1)}_{m}}\notag\\
&\qquad \qquad\qquad\qquad \times_{ k+1}{{\bf{A}}^{(k+1)}_{m}}\cdots\times_{N_{\rm {d}}}{\bf{A}}_m^{({N_{\rm d}})}.
\label{eq_35}
\end{align}

To facilitate the solution, we perform the $n$-mode unfolding of the tensor ${{\cal Y}_m}$ and ${\cal Z}_{m}$ in \eqref{eq_34}, as given by
\begin{align}
      \mathop  {\min }\limits_{{\left \{{{\bf{A}}^{(n)}_{m}}\right \}} } \quad  
&\lambda_1  \left\| {{{\bf Y}^{(n)}_{m}} - {{\bf{A}}^{(n)}_{m}}{{\bf Z}^{(n)}_{m}(\bigotimes^{N_{\rm {d}}}_{k\neq n} {{\bf{A}}^{(k)}_{m}})^{\mathsf {T}}  }} \right\|_F^2 \notag \\
&+\lambda_2\left \|  {{\bf{A}}^{(n)}_{m}} \right \|^2_F ,
\label{eq_36}
\end{align}
where ${{\bf Y}^{(n)}_{m}}$ and ${\bf Z}^{(n)}_{m}$ are the 
$n$-mode unfolded forms of ${{\cal Y}_m}$ and ${{\cal Z}_m}$; and
\begin{align}
&\bigotimes^{N_{\rm {d} }}_{n\neq k}{\bf {A}}^{(n)}_{m}={\bf {A}}^{(N_{\rm {d}})}_{m}\otimes\cdots\otimes {\bf {A}}^{(k+1)}_{m}\otimes \notag
\\ &\qquad\qquad\qquad {\bf {A}}^{(k-1)}_{m}\otimes\cdots\otimes {\bf {A}}^{(1)}_{m}.
\label{eq_37}
\end{align}

To enhance computational efficiency, we decompose the process of solving the factor matrix ${{{\bf{A}}^{(n)}_{m}}}$ into independently addressing each row, thereby simplifying the objective function as follows
\begin{align}
 \mathop{\min }\limits_{{{{\bf{a}}^{(n)}_{m,i^{(n)}_{\rm r}}}} } \quad  
&\lambda_1  \left\| {{{\bf y}^{(n)}_{m,o^{(n)}}} - {{{{\bf{a}}^{(n)}_{m,o^{(n)}}}} }{{\bf Z}^{(n)}_{m}(\bigotimes^{N_{\rm{d}}}_{k\neq n} {{\bf{A}}^{(k)}_{m}})^{\mathsf {T}}  }} \right\|_2^2 \notag \\
&+\lambda_2 \left \|  {{{{\bf{a}}^{(n)}_{m,o^{(n)}}}} } \right \|^2_2 ,
\label{eq_38}
\end{align}
where ${{{{\bf{a}}^{(n)}_{m,o^{(n)}_{}}}} }$ represents the ${o^{(n)}_{}}$-th row of ${{\bf{A}}^{(n)}_{m}}$, ${{\bf y}^{(n)}_{m,o^{(n)}_{}}}$ represents the ${o^{(n)}_{}}$-th row of ${{\bf Y}^{(n)}_{m}}$, $1\le {o^{(n)}_{}}\le {I}_n$, and ${I}_n$ represents the number of rows of ${{\bf Y}^{(n)}_{m}}$. Through the gradient method, the variable ${{{{\bf{a}}^{(n)}_{m,o^{(n)}_{}}}} }$ can be computed by
\begin{align}
{{{{\bf{a}}^{(n)}_{m,o^{(n)}_{}}}} }=\lambda_1{{{\bf y}^{(n)}_{m,o^{(n)}_{}}}} {\bf \Xi} ^{*}{\left ( \lambda_1 {\bf \Xi} ^{\sf T}{\bf \Xi} ^{\ast}+\lambda_2{\bf I}_\xi \right )^{-1}},
\label{eq_39}
\end{align}
where ${\bf I}_\xi$ is the identity matrix with the same dimension as 
\begin{align}
 {\bf \Xi}= \left( \mathop{\bigotimes^{N_{\rm{d}}}}\limits_{k \neq n} {\bf{A}}^{(k)}_{m} \right)({\bf{Z}}^{(n)}_{m})^{\mathsf {T}}.
 \label{eq_40}
\end{align}

\begin{algorithm}[t!]
\caption{Off-Grid NF Cascaded Channel Estimation Algorithm}
\label{alg:alg1}
\begin{algorithmic}[1] 
\REQUIRE  Received signal tensors $\left\{ \mathbf{\mathcal{Y}}_m \right\}_{m=1}^{M}$, 
regularization parameters $\lambda_1$, $\lambda_2$, logarithmic-adaptive parameter $\delta$, IRS phase shift $\mathbf{v}$.
\FOR{$m = 1$ to $M$}
    \STATE Initialize the core tensor $\mathcal{Z}_m^{[0]}$ and the factor matrix 
    $\left\{ \mathbf{A}_m^{(n)[0]} \right\}_{n=1}^{3}$ using \eqref{eq_44}--\eqref{eq_47}.
    \STATE Set $t = 1$.
    \WHILE{$t \leq t_{\max}$}
         \STATE Update  $\mathcal{Z}_m^{[t]}$ by using \eqref{eq_31}. 
           \FOR{$n = 1$ to $N_{\rm {d}}$}
                \FOR{$i_{\rm {r}} = 1$ to $I_{ {n}}$}
                 \STATE Update  ${{{{\bf{a}}^{(n)}_{m,i^{(n)}_{\rm r}}}} }$ using \eqref{eq_39}.
                \ENDFOR
           \ENDFOR
         \STATE $t = t + 1$.
    \ENDWHILE
   \STATE Obtain the estimated cascaded channel ${\hat{\cal{G}}}_m$ using \eqref{eq_41}.
\ENDFOR 
\ENSURE  Cascaded channel $\left \{ {\hat{\cal{G}}}_m \right \}_{m=1}^{M}  $.
\end{algorithmic}
\end{algorithm}

With Eqns. \eqref{eq_28}--\eqref{eq_33} and \eqref{eq_39}, we can obtain the core tensor ${\hat {\cal {Z}}_{m}^{}}$ and the estimated factor matrix ${\left \{{{\bf{\hat A}}^{(n)}_{m}}\right \}}$. The estimated cascaded channel at the $m$-th subcarrier is obtianed as
\begin{equation}
{{\hat{\bf{\cal G}}_m} = \left ( {\hat{\cal Z}_{m} \times_1{\hat{\bf{A}}}^{(1)}_{m}\times_2 {\hat{\bf{A}}}^{(2)}_{m}\times_3{\hat{\bf{A}}}^{(3)}_{m}} \right ) }\times_2{({\bf {V}}}^{\mathsf {T}})^{-1}.
\label{eq_41}
\end{equation}

The cascaded channel can be estimated through an iterative manner, where the choice of the initial values for the iterative estimation significantly affects both the accuracy and convergence speed of the algorithm. To address this, we employed the HOSVD method to initialize the factor matrices and core tensor required for the estimation process~\cite{r35}. Based on the HOSVD principle, the received tensor of the $m$-th subcarrier, ${{\bf{\cal Y}}_m}$, is decomposed as
\begin{equation}
{{{\bf{\cal Y}}_m} =  {{\cal L}_{m} \times_1{\bf{U}}^{(1)}_{m}\times_2 {\bf{U}}^{(2)}_{m}\times_3{\bf{U}}^{(3)}_{m}}  },
\label{eq_42}
\end{equation}
where ${\bf{U}}^{(i)}_{m}, i=1,2,3$ represents the left singular matrix of the mode-$n$ unfolding of the tensor ${{\bf{\cal Y}}_m}$. The core tensor ${\cal L}_{m}$ contains the structural information of the original tensor and can be expressed as
\begin{equation}
{{{\bf{\cal L}}_m} =  {{\cal Y}_{m} \times_1({\bf{U}}^{(1)}_{m})^{\mathsf {H}}\times_2 ({\bf{U}}^{(2)}_{m})^{\mathsf {H}}\times_3({\bf{U}}^{(3)}_{m}})^{\mathsf {H}}  }.
\label{eq_43}
\end{equation}

The HOSVD effectively captures the principal features of the original tensor ${{\bf{\cal Y}}_m}$ across each mode. Thus, the iterative solution for the initial values of the original core tensor and factor matrices are given by
\begin{align}
{{\cal Z}}^{[0]}_{m} &= {\cal L}_{m},
\label{eq_44}\\
({{\bf{A}}}^{(1)}_{m})^{[0]} &= {{\bf{U}}}^{(1)}_{m},
\label{eq_45}\\
({{\bf{A}}}^{(2)}_{m})^{[0]} &= {{\bf{U}}}^{(2)}_{m},
\label{eq_46}\\
\rm{and} \ ({{\bf{A}}}^{(3)}_{m})^{[0]} &= {{\bf{U}}}^{(3)}_{m},
\label{eq_47}
\end{align}
where $\left ( \cdot  \right ) ^{[0]}$ denotes the initial value of the iteration.

{Before optimization, the proposed method first normalizes the observed signal tensor based on the $\ell_2$-norm to improve numerical stability and reduce the impact of the original signal energy uncertainty on the algorithm. Specifically, the  $\ell_2$-norm of the observed tensor is computed, and the original tensor is divided by this norm to achieve energy normalization. Then, HOSVD is used to initialize the observed tensor, whose column dimensions directly determine the values of $G_{{\rm z}} $, $G_{{\rm r}}$ and $G_{{\rm y}}$. This method naturally endows the factor matrix columns with unit norm properties. As shown in \eqref{eq_45}--\eqref{eq_47}, the factor matrices obtained from HOSVD consist of the left singular vectors of the mode unfolding matrices. These vectors are not only of unit norm but also orthogonal to each other, thus constructing structurally stable subspace representations with normalized features across different modes, providing a numerically robust initialization point for the subsequent optimization process.
}{When the system employs  $M$ subcarriers, the proposed scheme requires solving the problem independently for each subcarrier. This design is primarily driven by the pronounced frequency selectivity of wideband mmWave communication, which necessitates subcarrier-dependent modeling to accurately capture the channel characteristics. Although our proposed method still requires solving problem (19) separately for each of the $M$ subcarriers, the algorithm significantly reduces the computational cost of each solve through sparse tensor decomposition and HOSVD-based dimensionality reduction, ensuring that the overall computational burden remains practical and competitive.}
The overall procedure of the above NF cascaded channel estimation method is summarized in Algorithm \ref{alg:alg1}.

{
\subsection{Multi-Antenna User Scenario}
In fact, our scheme can be easily extended to the multi-antenna user scenario. Assume that the user side is equipped with a ULA array consisting of ${N_{\rm ue}}$ antennas. Since the number of antennas at the user side is much smaller than that of the IRS, we assume that its array response follows the FF form \cite{r16}. The UE-IRS channel at the 
$m$-th subcarrier can then be expressed as
\begin{align}
{{\bf{H}}_{{\rm{u,}}m}} = & \sum\limits_{l = 1}^L {{\alpha_l}{e^{ - j2\pi {f_{m}}{\zeta_l}}}} {{\bf{a}}_{{\rm{tr}},m}}\left( {{\vartheta_{{\rm{e}},l}},{\varpi_{{\rm{a}},l}},{r_l}} \right) \notag \\
 & \times {{\bf{a}}_{{\rm{ue}},m}^{\mathsf{H}}}\left( {{\theta_{{\rm{ue}}}}{{{\rm{}}}}} \right) \in \mathbb{C}^{N_{\rm b}\times 1},
\label{r1}
\end{align}
where $L$ represents the number of paths; ${\alpha_l}$ and ${\zeta_l}$ are the complex path gain and time delay; ${\vartheta_{{\rm{e}},l}}$ and ${\varpi_{{\rm{a}},l}}$ 
are the elevation and azimuth AoAs; and ${r_l}$ is the distance between the IRS reference element and the UE.
 The expression of ${{\bf{a}}_{{\rm{tr}},m}}$  defined in a similar manner. The ${{\bf{a}}_{{\rm{ue}},m}^{\mathsf{}}}=[1,\cdots,e^{-j2\pi(N_{\rm{ue}} -1)\frac{f_m} {c}d{\lambda}\sin\theta_{\mathrm{ue}}}]^{{{\mathsf {T}}}} $ is the array response at the user side.
 During the uplink channel estimation phase, $T$ frames are used for training, and each frame consists of $P$ consecutive OFDM symbols. Let ${{\bf{s}}_{m,t,p}}\in \mathbb{C}^{N_{\rm ue}\times 1}$ denote the pilot sequence on the $m$-th subcarrier, in the $t$-th frame, and the $p$-th OFDM symbol. Then, the received signal ${{\bf y}_{m,t,p}}$ on the $m$-th subcarrier, in the $t$-th frame, and the $p$-th OFDM symbol is expressed as 
\begin{align}
{{\bf y}_{m,t,p}} = &{{\bf{H}}_{{\rm{g,}}m,t}}\operatorname{diag}({{\bf{v}}_p}){{\bf{H}}_{{\rm{u,}}m,t}}{\bf s}_{m,t,p} \notag \\&+ {{\bf n}_{m,t,p}}\in \mathbb{C}^{N_{\rm b}\times 1}.
\label{rr2}
\end{align}}

{
Vectorizing the \eqref{rr2} yields
\begin{align}
{{\bf y}_{m,t,p}}=&\left( {{\bf{v}}^{\mathsf T}_p}\otimes\mathbf{s}_{m,t,p}^{\mathsf T} \otimes \mathbf{I}_{N_{\text{b}}} \right)\operatorname{vec}({{\bf{H}}^{\mathsf T}_{{\rm{u,}}m,t}}\odot {{\bf{H}}_{{\rm{g,}}m,t}}) \notag \\&+ {{\bf n}_{m,t,p}}\in \mathbb{C}^{N_{\rm b}\times 1}.\label{rr3}
\end{align}
 } 
 
{
Collecting the signals during the $P$ OFDM symbols yields
\begin{align}
\mathbf{y}_{m,t} &= \left[ y_{m,t,1}^{\mathrm{T}}, \cdots, y_{m,t,P}^{\mathrm{T}} \right]^{\mathrm{T}} \notag\\
&= \left[ ({\bf V} \odot  {\bf S}_{m,t})^{\mathrm{T}} \otimes \mathbf{I}_{N_{\rm b}} \right] \operatorname{vec}({{\bf{H}}^{\mathsf T}_{{\rm{u,}}m,t}}\odot {{\bf{H}}_{{\rm{g,}}m,t}}) +{\bf n}_{m,t}  \notag\\
&= {\bf \Omega}_{m,t} {\bf u}_{m,t} + {\bf n}_{m,t} \in \mathbb{C}^{N_{\rm b}P \times 1},
\label{rr4}
\end{align}
where $\mathbf{V} = [{\bf v}_1, \cdots, {\bf v}_P] \in \mathbb{C}^{N_{\rm r} \times P}$, ${\bf S}_{m,t} = [s_{m,t,1}, \cdots, s_{m,t,P}] \in \mathbb{C}^{N_{\rm ue} \times P}$, ${\bf u}_{m,t}=\operatorname{vec}({{\bf{H}}^{\mathsf T}_{{\rm{u,}}m,t}}\odot {{\bf{H}}_{{\rm{g,}}m,t}})\in \mathbb{C}^{N_{\rm ue}N_{\rm r}N_{\rm b} \times 1}$, ${\bf n}_{m,t}=\left[ {\bf n}_{m,t,1}^{\mathrm{T}}, \cdots, {\bf n}_{m,t,P}^{\mathrm{T}} \right]^{\mathrm{T}}\in \mathbb{C}^{N_{\rm b}P \times 1}$.  Using least squares (LS), we can obtain the rough estimate of the channel ${\hat{\bf u}}_{m,t}={\bf \Omega}^{\dagger}_{m,t}\mathbf{y}_{m,t}$. The vector ${\bf u}_{m,t}$ can then be unvectorized to obtain 
\begin{align}
\mathbf{R}_{m,t} &\approx  \operatorname{unvec}({\hat {\bf{u}}}_{m,t})\notag\\
&\approx  {{\bf{H}}^{\mathsf T}_{{\rm{u,}}m,t}}\odot {{\bf{H}}_{{\rm{g,}}m,t}}\in \mathbb{C}^{N_{\rm ue}N_{\rm b} \times N_{\rm r}}\notag\\
&\approx   ( \mathbf{A}^{\ast}_{{\rm{ue}},m} \boldsymbol{\Lambda}^{\rm{a}}_{m,t} \mathbf{A}_{{\rm{tr}},m}^{\mathrm{T}}) \odot ( \mathbf{A}_{{\rm{b}},m} \boldsymbol{\Lambda}^{\rm{b}}_{m,t} \mathbf{A}_{{\rm{r}},m}^{\mathrm{H}})\notag\\
&\approx   ( \mathbf{A}^{\ast}_{{\rm{ue}},m} \otimes  \mathbf{A}_{{\rm{b}},m})\left [ \boldsymbol{\Lambda}^{\rm{a}}_{m,t}\otimes \boldsymbol{\Lambda}^{\rm{b}}_{m,t}   \right ] (\mathbf{A}_{{\rm{tr}},m}^{\mathrm{T}}\odot \mathbf{A}_{{\rm{r}},m}^{\mathrm{H}})\notag\\
 &\approx ( \mathbf{A}^{\ast}_{{\rm{ue}},m} \otimes  \mathbf{A}_{{\rm{b}},m})D_t({\bf F}_m){\bf B}_m^{\mathsf{T}}\in \mathbb{C}^{N_{\rm ue}N_{\rm b} \times N_{\rm r}},
 \label{rr5}
\end{align}
where $\mathbf{F}_{m} = [\mathbf{f}_{m,1}, \cdots, \mathbf{f}_{m,T}]^{\mathrm{T}} \in \mathbb{C}^{T \times L}$ with $ \mathbf{f}_{m,t} ={\boldsymbol{\alpha}}_{m,t} \otimes \boldsymbol{\beta}_{m,t} \in \mathbb{C}^{L \times 1}$. $D_t({\bf F}_m)$ is the diagonal matrix formed by the elements of the $t$-th row of matrix ${\bf F}_m$. ${\bf B}_m=(\mathbf{A}_{{\rm{tr}},m}^{\mathrm{T}}\odot \mathbf{A}_{{\rm{r}},m}^{\mathrm{H}})^{\mathsf{T}}\in \mathbb{C}^{{N_{\rm{r}}} \times L}$.}

{
By collecting $T$ time frames and exploiting the properties of the mode-$n$ product, we can obtain
\begin{align}
\mathcal{R}_{m} \approx \tilde{\mathcal{I}_m} \times_1 \tilde{ \mathbf{A}}_m \times_2 \mathbf{A}^{\ast}_{{\rm{ue}},m}  \times_3 {\bf B}_m ,
\label{rr6}
\end{align}
where $\tilde{ \mathbf{A}}_m=\mathbf{A}_{{\rm{b}},m}{\boldsymbol{\psi}_{1}}$.}

{
From \eqref{rr6}, it can be observed that each factor matrix is composed of the array response matrix at the BS, the array response matrix at the user side, and the incident-reflection array response matrix at the IRS. Meanwhile, the core tensor is constructed from the path gains and delay terms. This structure closely resembles that of \eqref{eq_11}. Therefore, by constructing the factor matrices in \eqref{rr6} as sparse transformation factor matrices using a method similar to that in the original work, and further expressing the core tensor in a sparse form, the proposed algorithm can be effectively applied to estimate the cascaded multi-antenna user channel.
 } 
{
  \begin{remark}
 Our proposed scheme can be further refined to reduce the pilot overhead by exploiting the Kronecker separable structure ~\cite{ad3} of the IRS phase-shift matrix. The prerequisites for utilizing a separable structure are readily satisfied in terahertz (THz) communications: given the severe path loss of THz channels, the non-line-of-sight (NLOS) components are typically negligible, making it reasonable to assume that the cascaded channel contains only a single dominant LOS path. Meanwhile, we consider decoupling the NF UPA response into a Kronecker-product form of two array responses. By further leveraging the Kronecker separable structure of the IRS phase-shift matrix, the overall cascaded channel estimation problem can be decomposed into two simplified subproblems, thereby significantly reducing the computational overhead ~\cite{ad3}.
\end{remark}
}
\section{Algorithm Analysis}\label{sec_4}
\subsection{Convergence Analysis}
We analyze the convergence of the proposed algorithm. The MM algorithm finds the solution by iteratively optimizing the surrogate function \eqref{eq_23} of the original problem. In each iteration, minimizing the surrogate objective ensures that the sequence of surrogate functions is non-increasing, i.e.,
\begin{align}
&S_{\rm {n}}\left({\cal{Z}}^{[t+1]}_m, \left\{\mathbf{A}_m^{(n)[t+1]}\right\} \mid {\cal{Z}}_m^{[t]}\right)\notag \\
 &\qquad \qquad\qquad \leq S_{\rm {n}}\left({\cal{Z}}^{[t]}_m, \left\{\mathbf{A}_m^{(n)[t]}\right\} \mid {\cal{Z}}_m^{[t]}\right).
 \label{eq_48}
\end{align}

Recalling the relationships among \eqref{eq_18}, \eqref{eq_19} and \eqref{eq_23}, we can arrive at the following inequality
\begin{align}
&L_{\rm o}\left({\cal{Z}}^{[t+1]}_m, \left\{(\mathbf{A}_m^{(n)})^{[t+1]}\right\}\right) \notag \\
&\qquad \qquad \qquad  - S_{\rm {n}}\left({\cal{Z}}^{[t+1]}_m, \left\{\mathbf{A}_m^{(n)[t+1]}\right\} \mid {\cal{Z}}_m^{[t]}\right) \notag\\
&=L_{\rm z}\left({\cal{Z}}^{[t+1]}_m\right) - f\left ( {{{\cal{Z}}}^{[t+1]}_{m}}  \mid {{{\cal{Z}}^{[t]}_m}} \right )\notag\\
&{\stackrel{(a)}\leq} L_{\rm z}\left({\cal{Z}}^{[t]}_m\right) - f\left ( {{{\cal{Z}}}^{[t]}_{m}}  \mid {{{\cal{Z}}^{[t]}_m}} \right ) \notag\\
&=L_{\rm o}\left({\cal{Z}}^{[t]}_m, \left\{(\mathbf{A}_m^{(n)})^{[t]}\right\}\right) - S_{\rm {n}}\left({\cal{Z}}^{[t]}_m, \left\{\mathbf{A}_m^{(n)[t]}\right\} \mid {\cal{Z}}_m^{[t]}\right), 
\label{eq_49}
\end{align}
where $(a)$ follows from the fact that the value of $f\left ( {{{\cal{Z}}}^{}_{m}}  \mid {{{\cal{Z}}^{[t]}_m}} \right )-L_{\rm z}\left({\cal{Z}}_m\right)$ is minimized when ${\cal{Z}}_m={\cal{Z}}^{[t]}_m$.

The objective for the ${[t+1]}$-th iteration is updated by
\begin{align}
&L_{\rm o}\left({\cal{Z}}^{[t+1]}_m, \left\{(\mathbf{A}_m^{(n)})^{[t+1]}\right\}\right)\notag\\
&=L_{\rm o}\left({\cal{Z}}^{[t+1]}_m, \left\{(\mathbf{A}_m^{(n)})^{[t+1]}\right\}\right)\notag \\
&\qquad  - S_{\rm {n}}\left({\cal{Z}}^{[t+1]}_m, \left\{\mathbf{A}_m^{(n)[t+1]}\right\} \mid {\cal{Z}}_m^{[t]}\right) \notag\\
&\qquad   +S_{\rm {n}}\left({\cal{Z}}^{[t+1]}_m, \left\{\mathbf{A}_m^{(n)[t+1]}\right\} \mid {\cal{Z}}_m^{[t]}\right). 
\label{eq_50}
\end{align}

Based on \eqref{eq_48}--\eqref{eq_50}, we have
\begin{align}
&L_{\rm o}\left({\cal{Z}}^{[t+1]}_m, \left\{(\mathbf{A}_m^{(n)})^{[t+1]}\right\}\right)\notag\\
&\qquad \le L_{\rm o}\left({\cal{Z}}^{[t]}_m, \left\{(\mathbf{A}_m^{(n)})^{[t]}\right\}\right) \notag\\
&\qquad \qquad - S_{\rm {n}}\left({\cal{Z}}^{[t]}_m, \left\{\mathbf{A}_m^{(n)[t]}\right\} \mid {\cal{Z}}_m^{[t]}\right) \notag\\
&\qquad \qquad  +S_{\rm {n}}\left({\cal{Z}}^{[t+1]}_m, \left\{\mathbf{A}_m^{(n)[t+1]}\right\} \mid {\cal{Z}}_m^{[t]}\right)\notag\\ 
&\qquad\le L_{\rm o}\left({\cal{Z}}^{[t]}_m, \left\{(\mathbf{A}_m^{(n)})^{[t]}\right\}\right).
\label{eq_51}
\end{align}
From \eqref{eq_51}, it is observed that the objective function is monotonically non-increasing with increasing iterations. Furthermore, we note that the objective in \eqref{eq_19} contains both log-sum terms and squared terms of the $F$-norm. As these terms approach zero, a lower bound for \eqref{eq_19} can be derived as
\begin{align}
L_{\rm o}\left({\cal{Z}}^{}_m, \left\{(\mathbf{A}_m^{(n)})^{}\right\}\right)\ge N_{\rm d} I_n  \log(\delta).
\label{eq_52}
\end{align}

Above all, since the objective function is monotonically decreasing and has a lower bound, the proposed algorithm is proven to be convergent.

\subsection{ Cram\'{e}r-Rao Bound Analysis}
Since CRLB can provide the lower bound for all unbiased estimators~\cite{r36}, it is widely adopted for performance analysis. 

{We first rewrite the form in (10) as 
    \begin{equation}
{{{\bf{\cal Y}}_m} = {{\bf{\cal G}}_m}\times_1{\bf {I}}_{N_{\rm z}}\times_2 {\bf{V}}^{\mathsf {T}}}\times_3{\bf {I}}_{N_{\rm y}}+{\cal N}_m \in \mathbb{C}^{N_{\rm z} \times P\times {N_{\rm y}}},
\label{eq_a1}
\end{equation}
where ${\bf {I}}_{N_{\rm z}} $ and ${\bf {I}}_{N_{\rm y}} $represent identity matrice of dimensions ${N_{\rm z}}$ and ${N_{\rm y}}$, respectively. Then we perform a mode-1 unfolding on ${{\bf{\cal Y}}_m}$ to obtain 
\begin{equation}
\mathcal{Y}_{m(1)} = \mathbf{I}_{N_{\rm z}} \mathcal{G}_{m(1)} \left( \mathbf{I}_{N_{\rm y}} \otimes \mathbf{V}^\mathsf{T} \right)^\mathsf{T} + \mathcal{N}_{m(1)}\in \mathbb{C}^{N_{\rm z} \times P {N_{\rm y}}},
\label{eq_a2}
\end{equation}
where $\mathcal{N}_{m(1)}$ represents the mode-1 unfolding of the noise tensor. Further, by vectorizing \eqref{eq_a2}, we have 
\begin{align}
{\bf{y}}_m &= {\rm {vec}}({{{\bf{\cal Y}}_{m(1)}}}) \in \mathbb{C}^{N_{\rm {p}} \times 1} \notag
\\&= \left( {\bf {I}}_{N_{\rm y}} \otimes {{\bf {V}}^{\mathsf {T}}} \otimes {\bf {I}}_{N_{\rm z}} \right) {\rm {vec}}({{{\bf{\cal G}}_{m(1)}}})+{\bf{n}}_m
\notag\\&={\bf {Q}}_{\rm v}{{\bf{h}}_m}+{\bf{n}}_m.
\label{eq_53}
\end{align}
where $N_{\rm {p}}=N_{\rm {z}}PN_{\rm {y}}$, ${{\bf h}_m}={\rm {vec}}({{{\bf{\cal G}}_{m(1)}}})\in \mathbb{C}^{{N_{\rm z}N_{\rm r}N_{\rm y}}\times 1}$, ${\bf {Q}}_{\rm v}=  {\bf {I}}_{N_{\rm y}} \otimes {{\bf {V}}^{\mathsf {T}}} \otimes {\bf {I}}_{N_{\rm z}}\in \mathbb{C}^{N_{\rm {p}} \times {N_{\rm z}N_{\rm r}N_{\rm y}}} $ and ${\bf{n}}_m={\rm {vec}}({{{\bf{\cal N}}_{m(1)}}})\in \mathbb{C}^{N_{\rm {p}} \times 1}$.
} 
%To derive the CRLB for the proposed channel estimator, we vectorize the received signal tensor ${{\bf{\cal Y}}_m}$ in \eqref{eq_10} as
%\begin{align}
%{\bf{y}}_m &= {\rm {vec}}({{{\bf{\cal Y}}_m}}) \in \mathbb{C}^{N_{\rm {p}} \times 1} \notag
%\\&= \left( {\bf {I}}_{N_{\rm z}} \otimes {{\bf {V}}^{\mathsf {T}}} \otimes {\bf {I}}_{N_{\rm y}} \right) {\rm {vec}}({{{\bf{\cal G}}_m}})+{\bf{n}}_m
%\notag\\&={\bf {Q}}_{\rm v}{{\bf{h}}_m}+{\bf{n}}_m, 
%\label{eq_53}
%\end{align}
%where $N_{\rm {p}}=N_{\rm {z}}PN_{\rm {y}}$, ${{\bf h}_m}={\rm {vec}}({{{\bf{\cal G}}_m}})\in \mathbb{C}^{{N_{\rm z}N_{\rm r}N_{\rm y}}\times 1}$, ${\bf {Q}}_{\rm v}=  {\bf {I}}_{N_{\rm z}} \otimes {{\bf {V}}^{\mathsf {T}}} \otimes {\bf {I}}_{N_{\rm y}}\in \mathbb{C}^{N_{\rm {p}} \times {N_{\rm z}N_{\rm r}N_{\rm y}}} $ and ${\bf{n}}_m={\rm {vec}}({{{\bf{\cal N}}_m}})\in \mathbb{C}^{N_{\rm {p}} \times 1}$.
Since the noise ${\bf{n}}_m$ follows a complex Gaussian distribution of $\mathcal{CN}(0,\sigma ^2 \mathbf{I}_{N_{\rm {p}} })$, the conditional probability density function of ${\bf{y}}_m $ given ${{\bf{h}}_m}$ can be written as
\begin{align}
p_{\mathbf{y}_m \mid \mathbf{h}_m}(\mathbf{y}_m; \mathbf{h}_m) = \frac{1}{(\pi\sigma^2)^{N_{\rm {p}}}} \exp \left\{ -\frac{1}{\sigma^2} \| \mathbf{y}_m - {\mathbf{Q}_{\rm {v}}}\mathbf{h}_m \|^2 \right\}.
\label{eq_54}
\end{align}
%By applying the natural logarithm to \eqref{eq_54}, we obtain:
%\begin{align}
%{L_{{\rm n},m}}&=\ln p_{\mathbf{y}_m \mid \mathbf{h}_m}(\mathbf{y}_m ; \mathbf{h}_m)\notag\\
%& =-{{N_{\rm {p}}}} \ln(\pi\sigma^2) 
%- \frac{1}{\sigma^2} \|\mathbf{y}_m - {\mathbf{Q}_{\rm {v}}}\mathbf{h}_m \|^2.
%\label{eq_55}
%\end{align}

The calculation of the CRLB requires the knowledge of the fisher information matrix (FIM)~\cite{r37}. The complex-valued FIM for ${{{\bf h}_m}}$ is given by
\begin{align}
{\bf{J}}\left( {{{\bf h}_m}} \right) &= \mathbb{E}\left\{ {{{\left( {\frac{\partial {L_{{\rm n},m}}}{{\partial {{\bf{h}}^{\sf T}_{{m}}}}}} \right)}^{\mathsf {H}}}\left( {\frac{\partial {L_{{\rm n},m}}}{{\partial {{\bf{h}}^{\sf T}_{{m}}}}}} \right)} \right\},
\label{eq_55}
\end{align}
{ 
      \noindent{}where
    \begin{align}
L_{n,m} &= \ln p_{\mathbf{y}_m \mid \mathbf{h}_m}(\mathbf{y}_m ; \mathbf{h}_m)\notag\\
&=\ln\left((\pi\sigma^2)^{-N_p}\right) + \ln\left(\exp\left\{ -\frac{1}{\sigma^2}\|{\mathbf{y}}_m - {\mathbf{Q}}_v{\mathbf{h}}_m\|^2 \right\}\right) \notag \\
&= -N_p\ln(\pi\sigma^2) - \frac{1}{\sigma^2}\|{\mathbf{y}}_m - {\mathbf{Q}}_v{\mathbf{h}}_m\|^2.
\label{eq_56}
\end{align}
} 
%\begin{align}
%{L_{{\rm n},m}}&=\ln p_{\mathbf{y}_m \mid \mathbf{h}_m}(\mathbf{y}_m ; \mathbf{h}_m)\notag\\
%& =-{{N_{\rm {p}}}} \ln(\pi\sigma^2) 
%- \frac{1}{\sigma^2} \|\mathbf{y}_m - {\mathbf{Q}_{\rm {v}}}\mathbf{h}_m \|^2.
%\label{eq_56}
%\end{align}
In \eqref{eq_55}, the partial derivative ${\frac{\partial {L_{{\rm n},m}}}{{\partial {{\bf{h}}^{\sf T}_{{m}}}}}}$ can be calculated as
{ 
 \begin{align}
\frac{\partial L_{n,m}}{\partial\mathbf{h}_m^{\mathsf {T}}}
&= -\frac{1}{\sigma^2}
\frac{\partial}{\partial\mathbf{h}_m^{\mathsf T}}
\Big(
- \mathbf{y}_m^{\mathsf H}\mathbf{Q}_v\mathbf{h}_m
- \mathbf{h}_m^{\mathsf H}\mathbf{Q}_v^{\mathsf H}\mathbf{y}_m\notag
\\& \ \ \ \ \ + \mathbf{h}_m^{\mathsf H}\mathbf{Q}_v^{\mathsf H}\mathbf{Q}_v\mathbf{h}_m
\Big)\notag\\
&= -\frac{1}{\sigma^2}
\left(-\mathbf{y}_m^{\mathsf H}\mathbf{Q}_v
+ \mathbf{h}_m^{\mathsf H}\mathbf{Q}_v^{\mathsf H}\mathbf{Q}_v \right) \notag\\
&= \frac{1}{\sigma^2}
\left(\mathbf{y}_m^{\mathsf H}\mathbf{Q}_v
- \mathbf{h}_m^{\mathsf H}\mathbf{Q}_v^{\mathsf H}\mathbf{Q}_v\right) \notag \\
 &=\frac{1}{\sigma^2}{\bf n}^{\mathsf {H}}_m\mathbf{Q}_{\rm v}.
 \label{eq_57}
\end{align}}
%\begin{align}
%{\frac{\partial {L_{{\rm n},m}}}{{\partial {{\bf{h}}^{\sf T}_{{m}}}}}} 
% &=\frac{-\frac{1}{\sigma^2} \left( - \mathbf{y}_m^{\mathsf {H}} \mathbf{Q}_{\rm v} \mathbf{h}_m - \mathbf{h}_m^{\mathsf {H}} \mathbf{Q}_{\rm v}^{\mathsf {H}} \mathbf{y}_m + \mathbf{h}_m^{\mathsf {H}} \mathbf{Q}_{\rm v}^{\mathsf {H}} \mathbf{Q}_{\rm v} \mathbf{h}_m \right)}{{\partial {{\bf{h}}^{\sf T}_{{m}}}}}\notag \\ 
% &=\frac{1}{\sigma^2}{\bf n}^{\mathsf {H}}_m\mathbf{Q}_{\rm v}.
 %\label{eq_57}
%\end{align}

To calculate the FIM, we first substitute \eqref{eq_57} into \eqref{eq_56}. Then, leveraging the linearity of the expectation operator and the Gaussian distribution of ${\mathbf{n}}_m$ yields

{ 
\begin{align}
\mathbf{J}(\mathbf{h}_m)
&= \mathbb{E}\left\{ \left(\frac{\partial L_{n,m}}{\partial \mathbf{h}_m^{{\mathsf T}}}\right)^{{\mathsf H}}
\left(\frac{\partial L_{n,m}}{\partial \mathbf{h}_m^{{\mathsf H}}}\right)\right\} \notag\\
&= \mathbb{E}\left\{ \left(\frac{1}{\sigma^2}\mathbf{n}_m^{{\mathsf H}}\mathbf{Q}_v\right)^{{\mathsf H}}
\left(\frac{1}{\sigma^2}\mathbf{n}_m^{{\mathsf H}}\mathbf{Q}_v\right)\right\}
 \notag\\
&= \frac{1}{\sigma^4}\,\mathbb{E}\left\{ \mathbf{Q}_v^{{\mathsf H}}\mathbf{n}_m\mathbf{n}_m^{{\mathsf H}}\mathbf{Q}_v \right\} \notag\\
&= \frac{1}{\sigma^4}\,\mathbf{Q}_v^{{\mathsf H}}\,\mathbb{E}\{\mathbf{n}_m\mathbf{n}_m^{{\mathsf H}}\}\,\mathbf{Q}_v \notag\\
&= \frac{1}{\sigma^4}\,\mathbf{Q}_v^{{\mathsf H}}\,(\sigma^2\mathbf{I}_{N_p})\,\mathbf{Q}_v \notag\\
&= \frac{1}{\sigma^2}\,\mathbf{Q}_v^{{\mathsf H}}\mathbf{Q}_v.
\label{eq_58}
\end{align}}
%\begin{align}
%{\bf{J}}\left( {{{\bf h}_m}} \right) 
%&= \frac{1}{\sigma^4}{\mathbb{E}}\left \{ \mathbf{Q}_{\rm{v}}^{\mathsf {H}} {{\bf n}_m}{{\bf n}^{\mathsf {H}}_m}\mathbf{Q}_{\rm{v}} \right \}  \notag \\
%&= \frac{1}{\sigma^2}\left \{ \mathbf{Q}_{\rm{v}}^{\mathsf {H}} \mathbf{Q}_{\rm{v}} \right \}.  
 %\label{eq_58}
%\end{align}

For the $m$-th subcarrier, the $\text{CRLB}_{m}$ for the channel ${{{\bf h}_m}}$ is obtained as
\begin{align}
\text{CRLB}_{m} &= \mathbb{E} \left\{ \left\| {\hat{\bf{h}}}_{m} - {\bf h}_{m} \right\|^2 \right\} \notag \\
&\geq \mathrm{Tr} \left\{ ({\mathbf{J}({\bf{h}}_m)})^{-1} \right\} = \sigma^2 \mathrm{Tr} \left\{ \left( \mathbf{Q}_{\rm{v}}^{\mathsf {H}} \mathbf{Q}_{\rm{v}} \right)^{-1} \right\}.
\label{eq_59}
\end{align}
From \eqref{eq_59}, it is observed that the CRLB is closely dependent on the phase shift of the IRS. To explore the specific relationship between them, $\left( \mathbf{Q}_{\rm{v}}^{\mathsf {H}} \mathbf{Q}_{\rm{v}} \right)^{-1}$ is expanded as
\begin{align}
    \left( {\bf {Q}}^{\mathsf {H}}_{\rm v} {\bf {Q}}_{\rm v} \right)^{-1} &= \left( {\left( {\bf {I}}_{N_{\rm y}} \otimes {{\bf {V}}^{\mathsf {T}}} \otimes {\bf {I}}_{N_{\rm z}}  \right)^{\mathsf {H}}} \left( {\bf {I}}_{N_{\rm y}} \otimes {{\bf {V}}^{\mathsf {T}}} \otimes {\bf {I}}_{N_{\rm z}} \right) \right)^{-1}\notag \\
 &= \mathbf{I}_{N_{\rm y}} \otimes (\mathbf{V}^* \mathbf{V}^{\mathsf {T}})^{-1} \otimes \mathbf{I}_{N_{\rm z}}.
 \label{eq_60}
\end{align}

Substituting \eqref{eq_60} into \eqref{eq_59}, the term $\mathrm{Tr} \left \{    \left( \mathbf{Q}_{\rm{v}}^{\mathsf {H}} \mathbf{Q}_{\rm{v}} \right)^{-1} \right \}$ becomes
\begin{align}
\mathrm{Tr} \left \{    \left( {\bf {Q}}^{\mathsf {H}}_{\rm v} {\bf {Q}}_{\rm v}  \right)^{-1} \right \}&=\mathrm{Tr} \left \{ \mathbf{I}_{N_{\rm y}} \otimes (\mathbf{V}^* \mathbf{V}^{\mathsf {T}})^{-1} \otimes \mathbf{I}_{N_{\rm z}}\right \}\notag\\
 &=\mathrm{Tr} \left \{ \mathbf{I}_{N_{\rm y}}\right \} \mathrm{Tr}\left \{ (\mathbf{V}^* \mathbf{V}^{\mathsf {T}})^{-1}\right \} \mathrm{Tr} \left \{  \mathbf{I}_{N_{\rm z}}\right \}\notag\\
&=N_{\rm {z}} N_{\rm {y}}  \operatorname{Tr}\left((\mathbf{V}^* \mathbf{V}^{\mathsf {T}})^{-1}\right). 
 \label{eq_61}
\end{align}

Based on \eqref{eq_61}, we have
\begin{align}
&\operatorname{Tr}\left((\mathbf{V}^* \mathbf{V}^{\mathsf {T}})^{-1}\right)= \sum_{{i_{\rm v}}=1}^{N_{\rm r}} \varrho _{i_{\rm v}}^{-1} \notag\\
&\qquad \qquad \overset{(a)}{\ge} {N_{\rm r}} \left( {{N_{\rm r}}}/{\sum\limits_{{i_{\rm v}}=1}^{N_{\rm r}}}\varrho _{i_{\rm v}} \right)=  \frac{{N_{\rm r}}^2}{\text{Tr} \left\{ \mathbf{V}^* \mathbf{V}^{\mathsf {T}} \right\}}. 
 \label{eq_62}
\end{align}
where $\varrho _{i_{\rm v}}$ denotes the ${i_{\rm v}}$-th singular value of $\mathbf{V}^* \mathbf{V}^{\mathsf {T}}$, and $(a)$ is derived from the arithmetic-harmonic mean inequality \cite{r38}. In inequality $(a)$, the equality holds if and only if all singular values $\left \{ \varrho _{i_{\rm v}} \right \}^{N_{\rm r}}_{{i_{\rm v}}=1} $ are equal. This indicates that the matrix $\mathbf{V}^{\mathsf {T}}$ is column-orthogonal. Since the elements of $\mathbf{V}$ are in the form of unit complex exponential functions, it can be concluded that $\mathbf{V}^* \mathbf{V}^{\mathsf {T}}$ is a diagonal matrix with diagonal elements equal to 
$P$, leading to $\text{Tr} \left\{ \mathbf{V}^* \mathbf{V}^{\mathsf {T}} \right\}={N_{\rm r}}P$.

Based on \eqref{eq_59}, \eqref{eq_61}, and \eqref{eq_62}, the CRLB for the 
$m$-th subcarrier can be expressed as
\begin{align}
\text{CRLB}_m=\sigma ^2\frac{N_{\rm z}N_{\rm y}N_{\rm r}}{P}. 
 \label{eq_63}
\end{align}
As a result, the CRLB under all subcarriers is given by
\begin{align}
\text{CRLB}=\sum_{m=1}^{M}\text{CRLB}_m=\sigma ^2\frac{M N_{\rm z}N_{\rm y}N_{\rm r}}{P}. 
 \label{eq_64}
\end{align}

\subsection{Complexity Analysis}
We discuss the computational complexity of the proposed channel estimiation approach. Based on Algorithm~\ref{alg:alg1}, the primary computational burden of the proposed method lies in computing the core tensor ${{{\cal{Z}}}_{m}}$ and the factor matrix $\left \{{{\bf{A}}^{(n)}_{m}}\right \}$ during each iteration.
According to \eqref{eq_43}, the complexity of performing HOSVD is given by ${\cal {O}}\left ( P(N_{\rm z}N_{\rm y})^2 \right )$.
By means of \eqref{eq_28}--\eqref{eq_33}, the complexity of calculating $ {{{\cal{Z}}}_{m}}$ is expressed as ${\cal {O}}\left ( G_{\rm z}G_{\rm r}G_{\rm y}N_{\rm z}+N_{\rm z}G_{\rm r}G_{\rm y}P+N_{\rm p}G_{\rm y}\right )$. Based on \eqref{eq_39}, the complexity of calculating the factor matrix is given by ${\cal {O}}( PN_{\rm b}G_{\rm z}G_{\rm r}G_{\rm y}+(G_{\rm z}^2+G_{\rm r}^2+G_{\rm y}^2)PN_{\rm b}$ $+G_{\rm z}^3N_{\rm z}+G_{\rm r}^3P+G_{\rm y}^3N_{\rm y}  )$. Therefore, the overall complexity of the proposed algorithm is expressed as${\cal{O}}( (PN_{\rm b}G_{\rm z}G_{\rm r}G_{\rm y}+(G_{\rm z}^2+G_{\rm r}^2+G_{\rm y}^2)PN_{\rm b}$ $+G_{\rm z}^3N_{\rm z}+G_{\rm r}^3P+G_{\rm y}^3N_{\rm y} +PN_{\rm r}^2 )M{t_{\max}})$. The benchmark algorithms are polar domain codebook-based NF IRS cascade channel estimation methods, including the block orthogonal least squares (P-BOLS) algorithm~\cite{r17} and the block orthogonal matching pursuit (P-BOMP) algorithm~\cite{r40}. The computational complexities of P-BOLS and P-BOMP are given by ${\cal {O}}\left ( L^3PMN_{\rm b}+ML^4+LMP^2N^2_{\rm b}N_G \right )$ and ${\cal {O}}\left ( L^3PMN_{\rm b}+ML^4+LPMN_{\rm b}N_G \right)$, respectively.  {The HDR~\cite{ad2} and the TSHDR~\cite{ad3} algorithms  are implemented within the HOSVD framework. These two algorithms have low complexity, but their estimation accuracy is inferior to that of our proposed scheme. }

%\begin{table}[!t]
%\captionsetup{font=small}
%\caption{Running Time of Different Methods\label{tab:table1}}
%\centering
%\begin{tabular}{|>{\centering\arraybackslash}m{0.1\textwidth}|>{\centering\arraybackslash}m{0.2\textwidth}|}
%\hline
%\textbf{Algorithm} & \textbf{Running Time (second)} \\ \hline
%P-BOMP  & 24.6369 \\ \hline
%P-BOLS  & 72.7193 \\ \hline
%Proposed  & 1.8888 \\ \hline
%\end{tabular}
%\end{table}

\section{Simulation Results} \label{sec_5}
This section presents numerical results to demonstrate the effectiveness of the proposed algorithm.
We consider an NF IRS-assisted broadband system, where a single-antenna UE transmits uplink pilot signals to the BS with the aid of the IRS. The path gains follow a complex Gaussian distribution, and the AoAs (AoDs) of all paths are uniformly generated from $\left( {0,2\pi } \right)$. {The NMSE is adopted to assess the algorithm accuracy, and its expression can be written as $\text{NMSE} =\frac{1}{{M}}  \sum_{m \in {M}} \mathbb{E} \left\{ \left\| {\hat{\bf{h}}}_{m} - {\bf h}_{m} \right\|^2/ \left\| {\bf h}_{m} \right\|^2\right\}
$~\cite{r39}}. { The comparison schemes adopted in this paper include the P-BOLS algorithm ~\cite{r17}, the T-OMP-JS~\cite{ad1}, the HDR algorithm ~\cite{ad2}, the TSHDR method ~\cite{ad3}, the P-BOMP algorithm ~\cite{r40}, and  }
Other simulation parameters are summarized in Table \ref{tab:table2}.

\begin{table}[!htbp]
%\captionsetup{font=small}
\caption{System Simulation Parameters\label{tab:table2}}
\centering
\begin{tabular}{|>{\centering\arraybackslash}m{0.32\textwidth}|>{\centering\arraybackslash}m{0.08\textwidth}|}
\hline
\textbf{Parameters} & \textbf{Value} \\ 
\hline 
Number of antennas for BS, $N_{\text{b}}$ & 25 \\ 
\hline
Number of antennas for IRS, $N_{\text{r}}$ & 256 \\ 
\hline
Carrier frequency $f$ & 28 GHz \\ 
\hline
Bandwidth $B$ & 2 GHz \\ 
\hline
Number of OFDM symbols, $P$ & 280 \\ 
\hline
Number of paths between the UE and the IRS, $L$ & 2 \\ 
\hline
{Regularization parameters $\lambda_1$ , $\lambda_2$ and $\delta$} & 
            {$\left [ 0.1,2 \right ] , 1$, $1e-10$} 
 \\ \hline
Number of subcarriers, $M$ & 6 \\ 
\hline
Distance between UE and IRS, $d_1$ & ${\cal{U}}(5,10)$\\ 
\hline
Distance between IRS and BS, $d_2$ & {$7.2153\,\text{m}$ }\\ 
\hline
\end{tabular}
\end{table}

%The benchmark algorithms are polar domain codebook-based NF IRS cascade channel estimation methods, including the block orthogonal least squares (P-BOLS) algorithm~\cite{r17} and the block orthogonal matching pursuit (P-BOMP) algorithm~\cite{r40}. 
%Additionally, the CRLB is used as a benchmark for comparison.
%Regularization parameters $\lambda_1$ and $\lambda_2$ & $\left [ 0.1,2 \right ] , 1$

\begin{figure}[!t]
\centering
\includegraphics[width=2.9in]{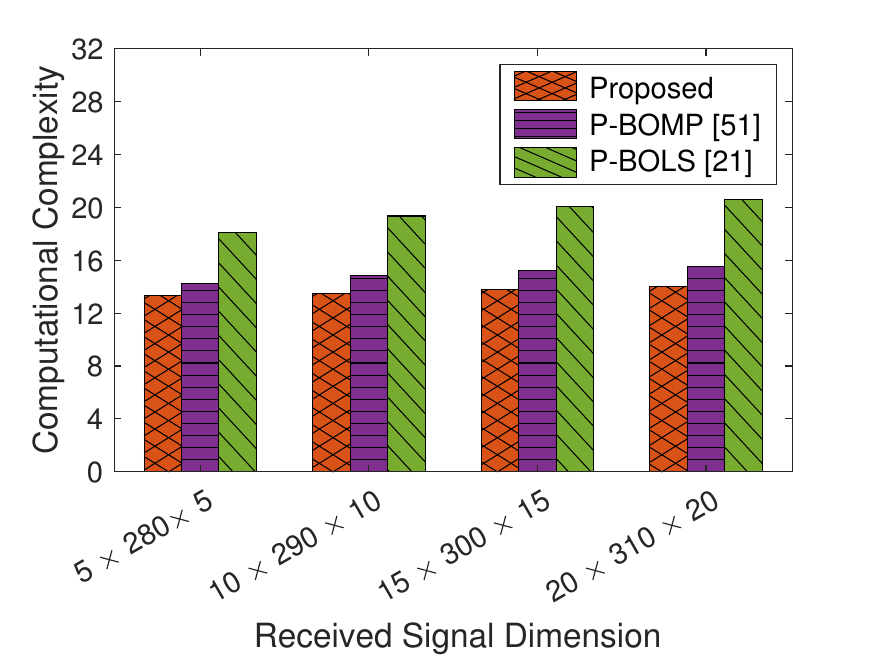}
\caption{Computational complexity comparison for different methods vs. dimension of the received signal. }
\label{fig_3}
\end{figure}

In Fig. \ref{fig_3}, we compare the computational complexity of the proposed algorithm and the benchmark algorithms. The $x$-axis represents the dimension of the received signal, while the $y$-axis represents the computational complexity in logarithmic scale (base 10). We set $G_{\rm{z}}=5$, $G_{\rm{r}}=280$, $t_{\max}=500$, $G_{\rm{y}}=5$, $G_{\rm{b}}=128$, $G_{\rm{u}}=128$  and $N_{\rm G} =G_{\rm b}G^{2}_{\rm u}$. The dimensions $N_{\rm{z}}$ and $N_{\rm{y}}$ of the received signal tensor vary from $5$ to $20$, and the value of 
$P$ varies from $280$ to $310$. It is seen that the computational complexity of the proposed algorithm is lower than that of the benchmark algorithms.
This is mainly because the CS methods based on polar domain codebooks generate a large sensing codebook for cascaded channel estimation. In contrast, the proposed approach avoids reliance on codebook searches, thereby significantly reducing algorithmic complexity.

\begin{figure}[!t]
\centering
\includegraphics[width=2.9in]{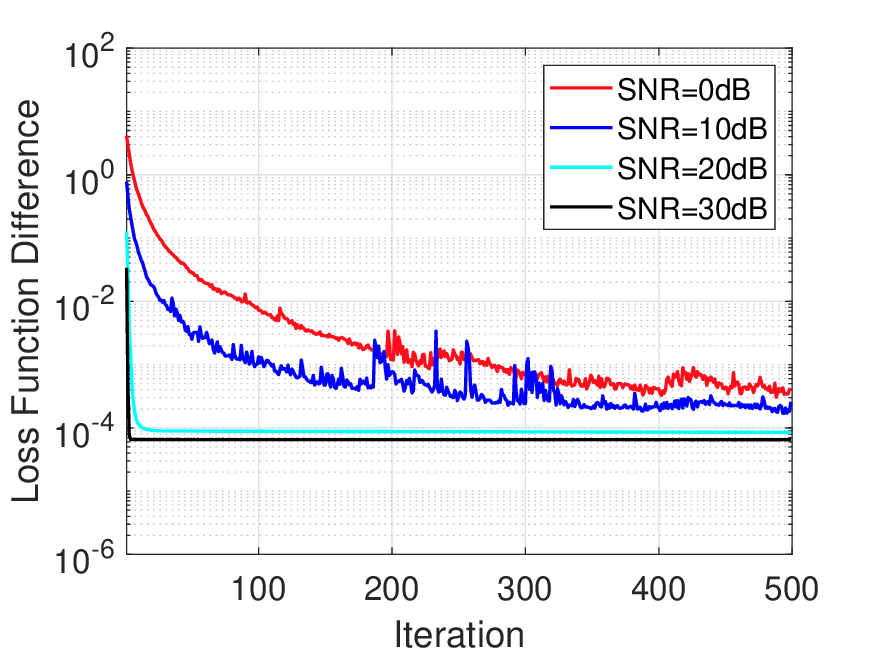}
\caption{Convergence behavior vs. SNR. }
\label{fig_4}
\end{figure}

Fig. \ref{fig_4} presents the convergence behaviors of the proposed algorithm across various SNR levels. As theoretically demonstrated in Section \ref{sec_4}-A, the convergence of the proposed algorithm is established. It can be observed that the convergence speed accelerates as the SNR increases.
When the SNR exceeds 10 dB, the algorithm rapidly converges within 20 iterations.

\begin{figure}[htbp]
\centering
\captionsetup[subfigure]{font=large} 
\subfloat[]{
    \includegraphics[width=2.9in]{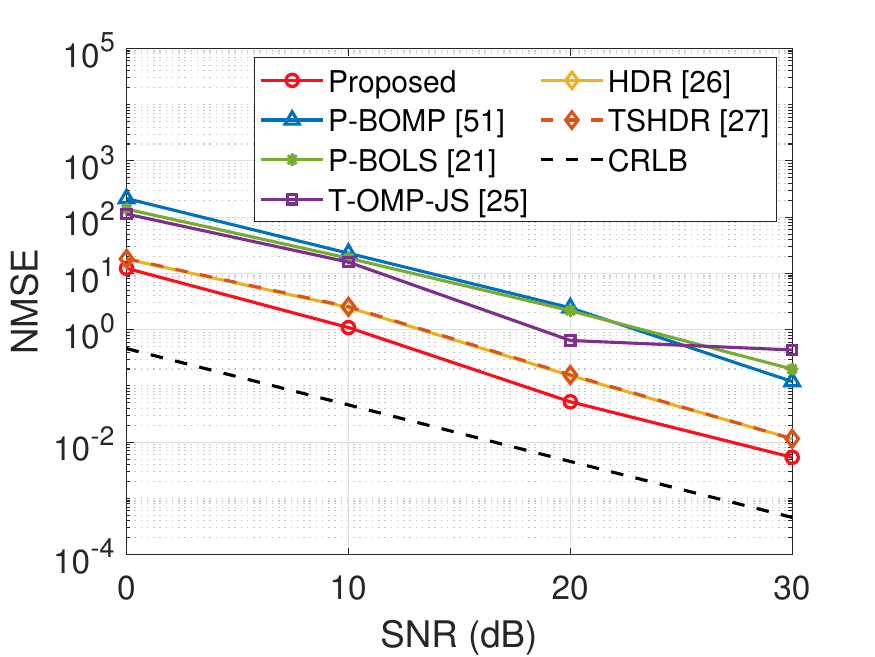}
    \label{fig_5a}
}
\hspace{0.1in}
\subfloat[]{
    \includegraphics[width=2.9in]{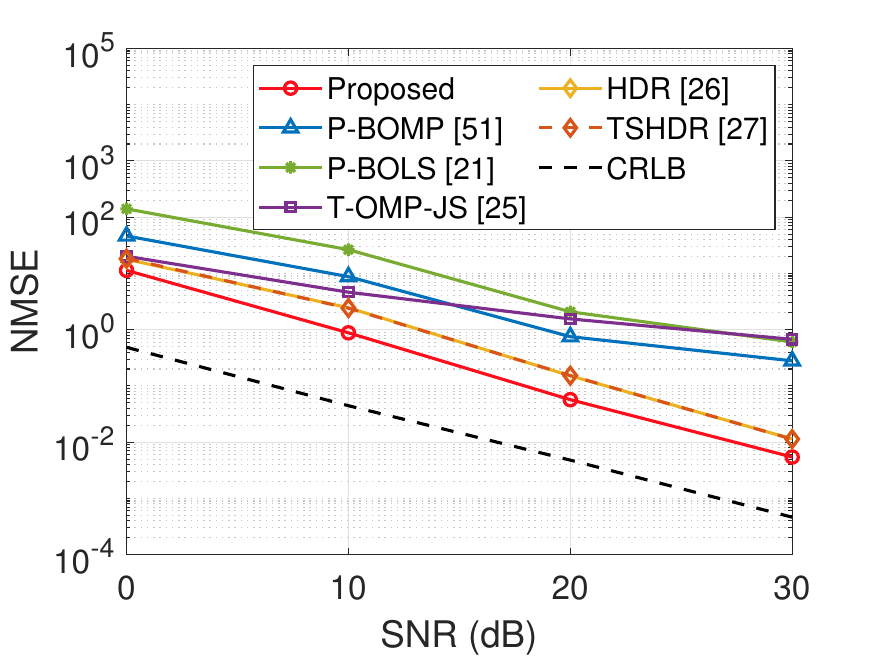}
    \label{fig_5b}
}
\caption{NMSE comparison for different algorithms. (a) UE-IRS distance of (5 m, 10 m). (b) UE-IRS distance of (10 m, 15 m).}
\label{fig_5}
\end{figure}

Fig.~\ref{fig_5} compares the NMSE performance curves of different algorithms. From Fig.~\ref{fig_5}(a), when the distance between the IRS and UE is set to (5 m, 10 m), the proposed algorithm consistently outperforms all benchmark methods across the entire SNR range. At low SNR levels, the P-BOMP and P-BOLS algorithms suffer from noise-induced selection errors and lack mechanisms for estimate refinement, resulting in significant performance degradation. In contrast, the proposed iterative approach operates in a continuous parameter space, allowing for progressive error correction and enhanced robustness to noise. 
Notably, the performance gap widens as SNR increases. The P-BOMP and P-BOLS methods experience saturation due to the resolution limitations of their discretized codebooks, while the proposed method continues to benefit from improved signal quality, showing a clear trend toward lower estimation error.  
Fig.~\ref{fig_5}(b) illustrates the NMSE performance under an extended UE-IRS distance of (10 m, 15 m). 
Consistent with Fig.~\ref{fig_5}(a), our algorithm maintains its performance advantage across all SNR levels. Despite the increased distance, our method delivers stable performance, whereas P-BOMP and P-BOLS methods exhibit significant performance fluctuations. This instability arises from the sensitivity of the polar-domain codebook to distance variations. 

{The T-OMP-JS algorithm capitalizes on the spatial structural sparsity of the cascaded channel. However, its estimation process depends on discrete codebook matrices for each mode and employs OMP to identify non-zero entries in the core tensor. This reliance, however, becomes its primary bottleneck: the final estimation accuracy is fundamentally capped by the resolution of the codebooks, thereby preventing the algorithm from achieving high-precision results. The HDR and TSHDR algorithms based on the PARAFAC tensor model fundamentally rely on the framework of HOSVD. However, HOSVD is essentially a global low-rank approximation method that captures only the overall low-rank structure of the tensor and cannot accurately exploit the inherent structured sparsity of the cascaded channel. Consequently, the estimation performance of these methods remains significantly inferior to that of our proposed scheme.}

\begin{figure}[!t]
\centering
\includegraphics[width=2.9in]{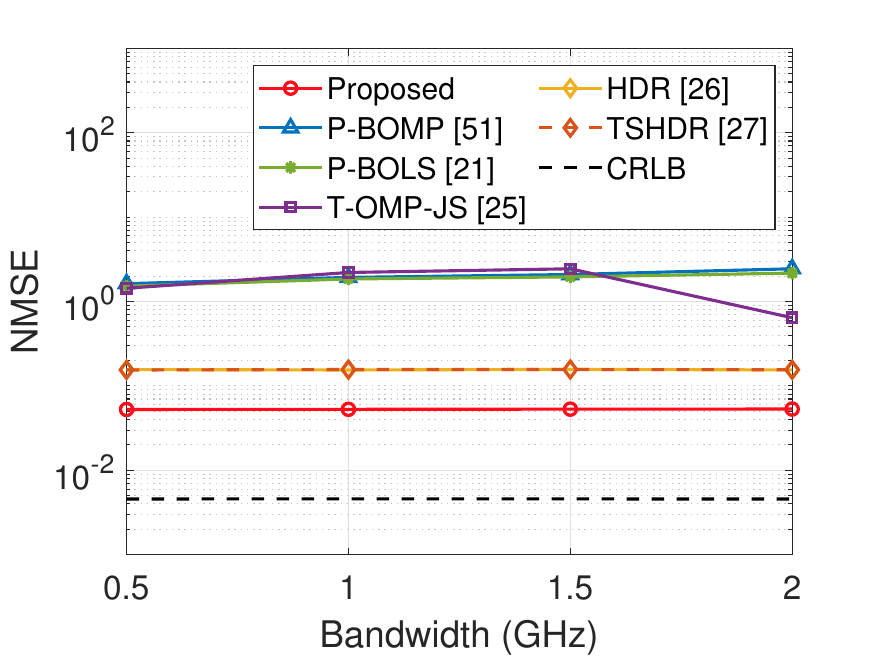}
\caption{NMSE comparison vs. system bandwidth at a SNR of 20 dB. }
\label{fig_7}
\end{figure}

\begin{figure}[!t]
\centering
\includegraphics[width=2.9in]{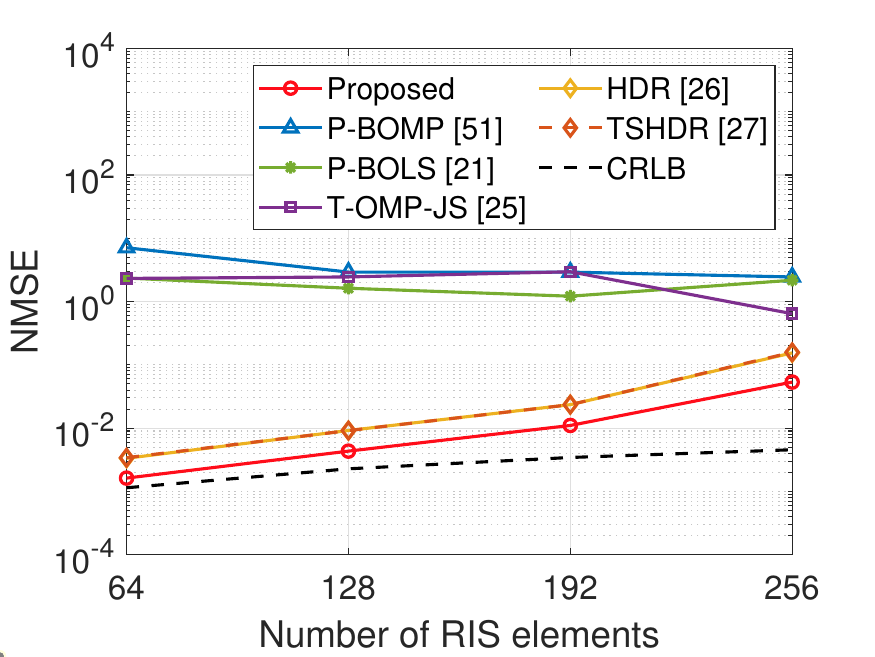}
\caption{NMSE comparison vs. number of IRS elements at a SNR of 20 dB. }
\label{fig_8}
\end{figure}

\begin{figure}[!t]
\centering
\includegraphics[width=2.9in]{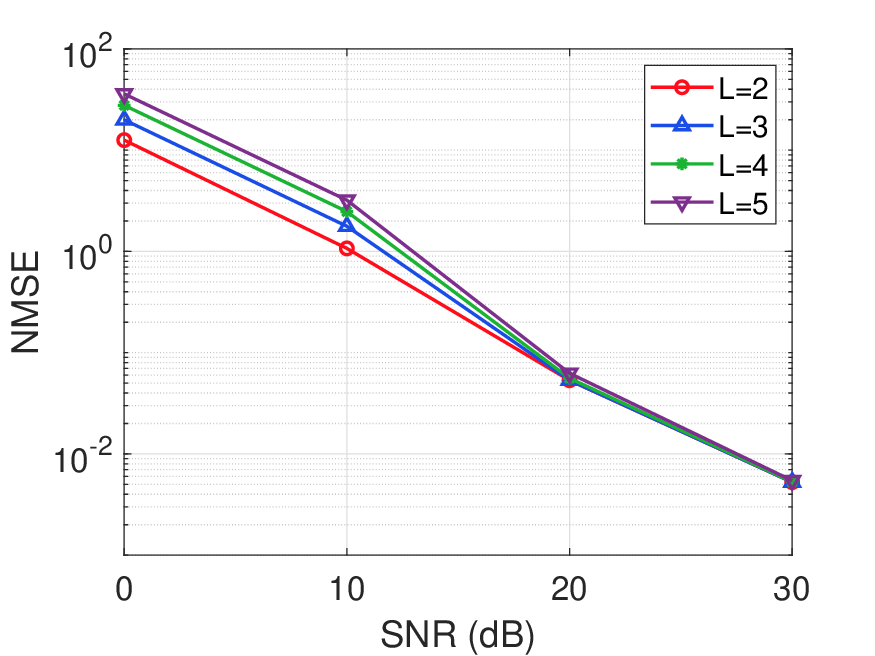}
\caption{NMSE comparison vs. IRS-UE path numbers. }
\label{fig_9}
\end{figure}

\begin{figure}[!t]
\centering
\includegraphics[width=2.9in]{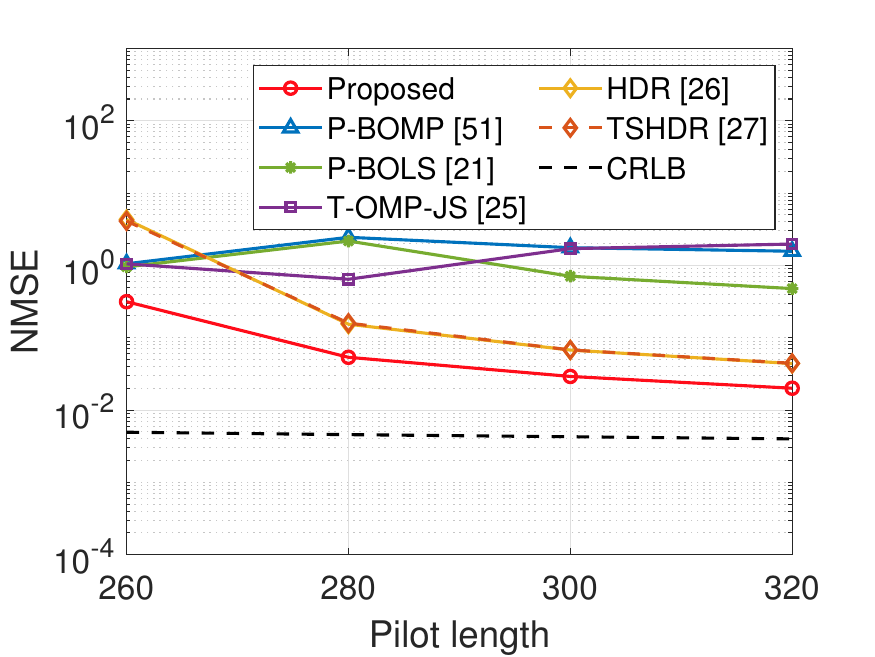}
\caption{NMSE comparison vs. pilot length at a SNR of 20 dB. }
\label{fig_10}
\end{figure}

In Fig. \ref{fig_7}, we compare the performance of different algorithms under various system bandwidths. We see that the proposed algorithm outperforms all comparison schemes in terms of NMSE performance, particularly under large bandwidth conditions. In contrast, the NMSE performance of the P-BOMP and P-BOLS schemes degrades progressively as bandwidth increases. This degradation stems from the frequency-dependent nature of the polar-domain codebook design. As bandwidth grows, the discrepancies among polar-domain codebooks at different frequencies become more significant, leading to increased channel estimation errors and reduced overall accuracy. In comparison, the proposed method achieves low and stable NMSE across the entire bandwidth range, demonstrating superior robustness and adaptability in wideband scenarios. {The T-OMP-JS algorithm exhibits unstable NMSE performance with varying bandwidth, primarily due to the random fluctuations in the degree of frequency-related codebook mismatch as bandwidth changes. The tensor-structured OMP algorithm amplifies the impact of these mismatches during the matching process, leading to overall performance instability. In contrast, the HDR and TSHDR algorithms, which do not use codebooks, demonstrate relatively stable estimation performance with varying bandwidth, but their accuracy is still inferior to that of the proposed scheme.}

Fig.~\ref{fig_8} compares the NMSE performance of the proposed algorithm with the benchmark algorithm as the number of IRS elements increases. While the NMSE of the comparison algorithms generally decreases with more IRS elements, the performance gains are ultimately limited by the quantization errors inherent in the polar-domain codebook, which constrain their estimation accuracy. In contrast, the NMSE performance of the proposed algorithm exhibits a slight degradation as the number of IRS elements increases. This behavior is primarily attributed to the increased condition number of the IRS-related matrix, which adversely affects the estimation due to matrix inversion in~\eqref{eq_41}. {In our system model, the HDR and TSHDR algorithms require a similar IRS matrix inversion process to remove its influence, which leads them to exhibit a trend similar to that of our proposed scheme. However, their overall estimation performance still shows a clear gap compared with our approach. For the T-OMP-JS algorithm, variations in the number of IRS elements alter the row dimension of its polar-domain codebook factor matrix. Random fluctuations of grid mismatch arise in this dimension due to the nonlinear characteristics of the angle-distance parameters in the polar-domain codebook, and these fluctuations propagate to all dimensions through the multi-dimensional coupling of the tensor, ultimately resulting in unstable NMSE performance.} Nevertheless, the proposed off-grid estimation method consistently outperforms the benchmark algorithms across all configurations, demonstrating superior robustness and effectiveness as the complexity of the IRS system scales.

As illustrated in Fig. \ref{fig_9}, the NMSE performance of the proposed algorithm is evaluated under varying numbers of IRS-UE propagation paths across different SNR levels. The results show that when the SNR is below 20 dB, an increase in the number of paths between the IRS and the UE reduces the sparsity of the core tensor, thereby amplifying the impact of noise and leading to degraded NMSE performance. In contrast, at higher SNR levels, the influence of noise is significantly mitigated, enabling more accurate signal recovery. In this regime, the variation in the number of IRS-UE paths has minimal effect on NMSE performance, highlighting the robustness of the proposed method under favorable SNR conditions.

Fig.~\ref{fig_10} evaluates the NMSE performance as the pilot length increases.
The proposed algorithm consistently benefits from extended pilot sequences, effectively exploiting the additional pilot information to enhance estimation accuracy. {Similarly, the HDR and TSHDR methods can also effectively utilize pilot information, but their estimation performance still lags behind that of our proposed scheme. The performance of the T-OMP-JS, P-BOMP, and P-BOLS algorithms becomes increasingly unstable as the pilot length increases. This performance fluctuation primarily arises from their limited ability to handle high-dimensional information. Therefore, even with an increase in pilot length, these on-grid methods fail to achieve significant performance improvement.} 

\begin{figure}[!t]
\centering
\includegraphics[width=2.9in]{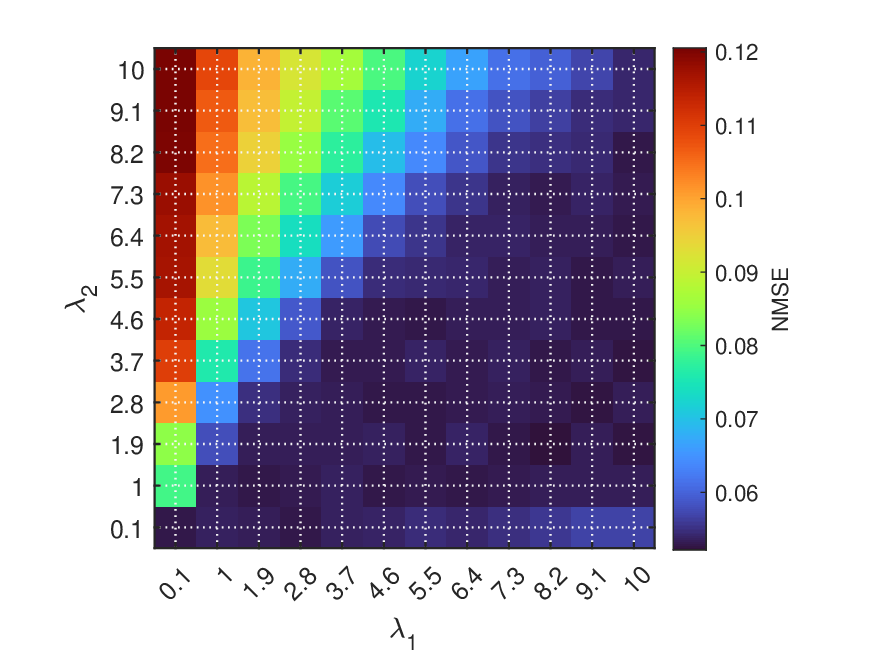}
\caption{{Heatmap of NMSE versus $\lambda_1$ and $\lambda_2$ for the proposed scheme at a SNR of 20 dB.}}
\label{fig_11}
\end{figure}

%\begin{figure}[!t]
%\centering
%\includegraphics[width=2.9in]{eps3/D_e.eps}
%\caption{\textcolor{blue}{NMSE of the proposed scheme versus $\delta$.}}
%\label{fig_12}
%\end{figure}

{ Fig.~\ref{fig_11} presents the NMSE heatmap of the proposed algorithm as a function of $\lambda_1$ and  $\lambda_2$. The results show that the algorithm maintains stable reconstruction accuracy over a wide range of parameters, while noticeable performance degradation concentrates in the region where $\lambda_1$ is too small and $\lambda_2$ is too large. Mechanistically, $\lambda_1$ weights the data-fidelity term; when it is too low, the constraint on reconstruction error is weakened, leading to underfitting. Meanwhile, a large $\lambda_2$ imposes strong shrinkage on the factor matrices, limiting their expressive capacity; coupled with the non-convex nature of the log-sum sparsity prior, this makes incorrect sparse-support selection more likely, thereby raising NMSE. In contrast, across a broad region with moderate $\lambda_1$ and small-to-moderate $\lambda_2$, the data term and regularization are well balanced, enabling stable and accurate recovery of the channel structure and rendering NMSE relatively insensitive to parameter variations (i.e., the surface changes smoothly). Overall, this sensitivity analysis indicates that the method’s performance does not hinge on a particular parameter setting and exhibits strong robustness over a wide parameter range.}

%\textcolor{blue}{
  %  We performed a wide-range sensitivity test on $\delta$  and observed only marginal NMSE variations across different $\delta$  values, as shown in Fig.~\ref{fig_12}. This is because the nonzero entries of the core tensor correspond to effective propagation paths in the near-field cascaded mmWave channel, whose path gains and delay-related parameters exhibit pronounced physical magnitudes; meanwhile, mmWave propagation features strong energy concentration over a few dominant paths, making these nonzero coefficients typically far larger than $\delta$. Therefore, $\delta$ has a negligible impact on sparsity modeling and estimation accuracy, and mainly serves as a numerical safeguard without requiring delicate tuning.}

\section{Conclusion} \label{sec_6}
This paper addressed the challenges of \textit{codebook dimension explosion} and \textit{quantization error accumulation} in NF cascaded channel estimation for UPA-architecture BS-IRS systems in 6G communications. 
We proposed a tensor modelization approach that substantially reduced codebook overhead while streamlining the estimation process.
Additionally, we developed an off-grid estimation framework based on sparse Tucker tensor decomposition, where a sparse core tensor minimization problem with tri-modal log-sum sparse constraints was constructed. By combining HOSVD preprocessing, the MM method, and the TOMFISTA technique, our approach achieved superior accuracy and efficiency. Theoretical analysis established the conditions for convergence to the CRLB and rigorously proved the algorithm convergence. Simulation results confirmed that the proposed method surpassed existing methods in terms of NMSE and computational efficiency, offering a promising solution for XL-IRS systems. 

%\appendices
%\section{Derivation of the FIM Matrix}
%\label{appendix:FIM}

\bibliographystyle{IEEEtran}
\bibliography{myref}

\begin{thebibliography}{10}
\providecommand{\url}[1]{#1}
\csname url@samestyle\endcsname
\providecommand{\newblock}{\relax}
\providecommand{\bibinfo}[2]{#2}
\providecommand{\BIBentrySTDinterwordspacing}{\spaceskip=0pt\relax}
\providecommand{\BIBentryALTinterwordstretchfactor}{4}
\providecommand{\BIBentryALTinterwordspacing}{\spaceskip=\fontdimen2\font plus
\BIBentryALTinterwordstretchfactor\fontdimen3\font minus \fontdimen4\font\relax}
\providecommand{\BIBforeignlanguage}[2]{{%
\expandafter\ifx\csname l@#1\endcsname\relax
\typeout{** WARNING: IEEEtran.bst: No hyphenation pattern has been}%
\typeout{** loaded for the language `#1'. Using the pattern for}%
\typeout{** the default language instead.}%
\else
\language=\csname l@#1\endcsname
\fi
#2}}
\providecommand{\BIBdecl}{\relax}
\BIBdecl

\bibitem{r1}
Z.~Chen, G.~Chen, J.~Tang, S.~Zhang, D.~K.~C. So, O.~A. Dobre, K.-K. Wong, and J.~Chambers, ``Reconfigurable-intelligent-surface-assisted {B5G/6G }wireless communications: {C}hallenges, {S}olution, and {F}uture {O}pportunities,'' \emph{IEEE Commun. Mag.}, vol.~61, no.~1, pp. 16--22, Sep. 2023.

\bibitem{r2}
T.~Zhang and S.~Mao, ``Joint beamforming design in reconfigurable intelligent surface-assisted rate splitting networks,'' \emph{IEEE Trans. Wireless Commun.}, vol.~23, no.~1, pp. 263--275, May 2024.

\bibitem{r3}
M.~M. Amri, N.~M. Tran, J.~H. Park, D.~I. Kim, and K.~W. Choi, ``Sparsity-aware channel estimation for fully passive {RIS}-based wireless communications: {T}heory to experiments,'' \emph{IEEE Internet Things J.}, vol.~10, no.~9, pp. 8046--8067, Dec. 2023.

\bibitem{r4}
R.~Schroeder, J.~He, G.~Brante, and M.~Juntti, ``Two-stage channel estimation for hybrid {RIS} assisted mimo systems,'' \emph{IEEE Trans. Commun.}, vol.~70, no.~7, pp. 4793--4806, May 2022.

\bibitem{r5}
J.~He, M.~Niu, P.~Zhang, and C.~Qin, ``Enhancing {PHY}-layer authentication in {RIS}-assisted {IoT} systems with cascaded channel features,'' \emph{IEEE Internet Things J.}, vol.~11, no.~14, pp. 24\,984--24\,997, Apr. 2024.

\bibitem{r6}
K.~Li, C.~Huang, Y.~Gong, and G.~Chen, ``Double deep learning for joint phase-shift and beamforming based on cascaded channels in {RIS}-assisted {MIMO} networks,'' \emph{IEEE Wireless Commun. Lett.}, vol.~12, no.~4, pp. 659--663, Jan. 2023.

\bibitem{r7}
J.~Liu, G.~Yang, Y.~Liu, and X.~Zhou, ``{RIS} empowered near-field covert communications,'' \emph{IEEE Trans. Wireless Commun.}, vol.~23, no.~10, pp. 15\,477--15\,492, Jul. 2024.

\bibitem{r8}
X.~Mu, J.~Xu, Y.~Liu, and L.~Hanzo, ``Reconfigurable intelligent surface-aided near-field communications for 6{G}: opportunities and challenges,'' \emph{IEEE Veh. Technol. Mag.}, vol.~19, no.~1, pp. 65--74, Jan. 2024.

\bibitem{r9}
J.~Wang, J.~Xiao, Y.~Zou, W.~Xie, and Y.~Liu, ``Wideband beamforming for {RIS} assisted near-field communications,'' \emph{IEEE Trans. Wireless Commun.}, vol.~23, no.~11, pp. 16\,836--16\,851, Dec. 2024.

\bibitem{r10}
T.~A. D.~W. M.~Guerra and E.~Hossain, ``Channel estimation in {RIS}-aided mmwave wireless systems using matching pursuit with phase rotation,'' \emph{IEEE Trans. Wireless Commun.}, vol.~23, no.~10, pp. 13\,187--13\,201, May 2024.

\bibitem{r11}
H.~Yang, A.~Zhang, Y.~Sun, J.~Li, and P.~Liu, ``Regularized sparse bayesian learning based channel estimation for {RIS}-assisted wireless communication system,'' \emph{IEEE Commun. Lett.}, vol.~28, no.~6, pp. 1412--1416, Mar. 2024.

\bibitem{r12}
A.~Abdallah, A.~Celik, M.~M. Mansour, and A.~M. Eltawil, ``{RIS}-aided mmwave {MIMO} channel estimation using deep learning and compressive sensing,'' \emph{IEEE Trans. Wireless Commun.}, vol.~22, no.~5, pp. 3503--3521, Mar. 2023.

\bibitem{r13}
J.~Chen, Y.-C. Liang, H.~V. Cheng, and W.~Yu, ``Channel estimation for reconfigurable intelligent surface aided multi-user mmwave {MIMO} systems,'' \emph{IEEE Trans. Wireless Commun.}, vol.~22, no.~10, pp. 6853--6869, Feb. 2023.

\bibitem{a}
H.~Sun, L.~Zhu, W.~Mei, and R.~Zhang, ``Power measurement-based channel estimation for {IRS}-enhanced wireless coverage,'' \emph{IEEE Trans. Wirel. Commun.}, vol.~23, no.~12, pp. 19\,183--19\,198, Dec. 2024.

\bibitem{bbb}
------, ``Power-measurement-based channel autocorrelation estimation for {IRS}-assisted wideband communications,'' \emph{IEEE Trans. Wirel. Commun.}, vol.~24, no.~6, pp. 4647--4662, Jun. 2025.

\bibitem{Re2-16}
X.~Zheng, P.~Wang, J.~Fang, and H.~Li, ``Compressed channel estimation for irs-assisted millimeter wave ofdm systems: A low-rank tensor decomposition-based approach,'' \emph{IEEE Wirel. Commun. Lett.}, vol.~11, no.~6, pp. 1258--1262, Jun. 2022.

\bibitem{C-Z-21}
C.~Zhou, C.~You, S.~Gong, B.~Lyu, B.~Zheng, and Y.~Gong, ``Channel estimation for {XL-IRS} assisted wireless systems with double-sided visibility regions,'' in \emph{Proc. Int. Conf. on Wireless Commun. and Signal Process. (WCSP)}.\hskip 1em plus 0.5em minus 0.4em\relax Hefei, China: IEEE, 2024, pp. 456--461.

\bibitem{r14}
Z.~Tang, Y.~Chen, Y.~Wang, T.~Mao, Q.~Wu, M.~D. Renzo, and L.~Hanzo, ``Near-field sparse channel estimation for extremely large-scale {RIS}-aided wireless communications,'' in \emph{Proc. IEEE Globecom. Workshops (GC Wkshps)}, Kuala Lumpur, Malaysia, Mar. 2023, pp. 1373--1379.

\bibitem{r15}
X.~Yu, W.~Shen, R.~Zhang, C.~Xing, and T.~Q.~S. Quek, ``Channel estimation for {XL-RIS}-aided millimeter-wave systems,'' \emph{IEEE Trans. Commun.}, vol.~71, no.~9, pp. 5519--5533, Jun. 2023.

\bibitem{r16}
S.~Yang, W.~Lyu, Z.~Hu, Z.~Zhang, and C.~Yuen, ``Channel estimation for near-field {XL-RIS}-aided mmwave hybrid beamforming architectures,'' \emph{IEEE Trans. Veh. Technol.}, vol.~72, no.~8, pp. 11\,029--11\,034, Mar. 2023.

\bibitem{r17}
J.~Wu, S.~Kim, and B.~Shim, ``Parametric sparse channel estimation for {RIS}-assisted terahertz systems,'' \emph{IEEE Trans. Commun.}, vol.~71, no.~9, pp. 5503--5518, Jun. 2023.

\bibitem{r18}
S.~Yang, C.~Xie, W.~Lyu, B.~Ning, Z.~Zhang, and C.~Yuen, ``Near-field channel estimation for extremely large-scale reconfigurable intelligent surface {XL-RIS}-aided wideband mmwave systems,'' \emph{IEEE J. Sel. Areas Commun.}, vol.~42, no.~6, pp. 1567--1582, Mar. 2024.

\bibitem{ad0}
S.~Cheng, L.~You, Z.~Jin, L.~Cheng, and X.~Gao, ``Tensor-based channel estimation for near-field millimeter wave {XL-MIMO} systems,'' \emph{IEEE Wireless Commun. Lett.}, 2025, early access, doi: 10.1109/LWC.2025.3558907.

\bibitem{t1}
X.~Guo, Z.~Xie, J.~Zhang, S.~Chen, and C.~Zhu, ``Beamspace channel estimation via {PARAFAC} decomposition for {RIS} assisted millimeter-wave multiuser {MISO} communications,'' \emph{IEEE Trans. Veh. Technol.}, pp. 1--15, Dec. 2024.

\bibitem{ad1}
D.~C. Ara´ujo, A.~L.~F. de~Almeida, J.~P. C. L.~D. Costa, and R.~T. de~Sousa, ``Tensor-based channel estimation for massive {MIMO-OFDM} systems,'' \emph{IEEE Access}, vol.~7, pp. 42\,133--42\,147, Mar. 2019.

\bibitem{ad2}
F.~E. Asim, B.~Sokal, A.~L.~F. de~Almeida, B.~Makki, and G.~Fodor, ``Structured channel estimation for {RIS}-assisted {TH}z communications,'' \emph{IEEE Trans. Veh. Technol.}, vol.~74, no.~3, pp. 5175--5180, Mar. 2025.

\bibitem{ad3}
F.~E. Asim, A.~L.~F. de~Almeida, B.~Sokal, B.~Makki, and G.~Fodor, ``Two-dimensional channel parameter estimation for {IRS}-assisted networks,'' \emph{IEEE Trans. Commun.}, vol.~73, no.~8, pp. 6337--6350, Aug. 2025.

\bibitem{r20}
X.~Xu, S.~Zhang, F.~Gao, and J.~Wang, ``Sparse bayesian learning based channel extrapolation for {RIS} assisted {MIMO-OFDM},'' \emph{IEEE Trans. Commun.}, vol.~70, no.~8, pp. 5498--5513, Jun. 2022.

\bibitem{r21}
Y.~Shi, Y.~Huang, X.~W. Tang, and Y.~Xiao, ``Channel estimation for wideband mmwave {MIMO-OFDM} system with beam squint effect,'' \emph{IEEE Commun. Lett.}, vol.~28, no.~1, pp. 153--157, Dec. 2024.

\bibitem{r22}
S.~Lv, Y.~Liu, X.~Xu, A.~Nallanathan, and A.~L. Swindlehurst, ``{RIS}-aided near-field {MIMO} communications: codebook and beam training design,'' \emph{IEEE Trans. Wireless Commun.}, vol.~23, no.~9, pp. 12\,531--12\,546, May 2024.

\bibitem{r23}
S.~Tang, Z.~Dong, W.~Zhou, D.~Cai, E.~Zhou, and P.~Lan, ``Beam splitting sensing for terahertz near-field channel estimation,'' in \emph{Proc. Int. Conf. Intell. Commun. Comput. (ICC)}, Nanchang, China, Nov. 2023, pp. 141--146.

\bibitem{r24}
H.~Ozen, O.~Yilmaz, and G.~M. Guvensen, ``Beam-squint-aware channel estimation for dual-wideband {UPA}-type {RIS}-aided massive {MIMO},'' \emph{IEEE Commun. Lett.}, vol.~28, no.~10, pp. 2367--2371, Aug. 2024.

\bibitem{r25}
L.~Chen, W.~Zhu, T.~Zhang, K.~Qiao, Y.~Jia, and Q.~Leng, ``Multi-path-based cascaded channel estimation for reconfigurable intelligent surface assisted multi user {MIMO }system,'' in \emph{Proc. Int. Conf. Commun. Technol. Inf. Technol. (ICCTIT)}, Xi'an, China, Feb. 2023, pp. 88--91.

\bibitem{G1}
Y.~Liu, J.~Xu, Z.~Wang, X.~Mu, and L.~Hanzo, ``Near-field communications: What will be different?'' \emph{IEEE Wireless Commun.}, vol.~32, no.~2, pp. 262--270, Mar. 2025.

\bibitem{G2}
W.~Long, W.~Song, Y.~Liu, Y.~Liu, M.~Moretti, and R.~Chen, ``{GPS}-denied {ISAC} vehicle localization based on mmwave radar and identification,'' \emph{IEEE Open J. Veh. Technol.}, vol.~6, pp. 2343--2357, Aug. 2025.

\bibitem{r26}
T.~G. Kolda, ``\textit{Multilinear operators for higher-order decompositions},'' Sandia National Laboratories, Tech. Rep. SAND2006-2081, Apr 2006.

\bibitem{Tda}
T.~G. Kolda and B.~W. Bader, ``Tensor decompositions and applications,'' \emph{SIAM review}, vol.~51, no.~3, pp. 455--500, Aug. 2009.

\bibitem{Ra}
M.~Liu, T.~Lin, Y.~Zhu, and Y.-J.~A. Zhang, ``Sparse channel estimation for {IRS}-assisted millimeter wave {MIMO}-{OFDM} systems,'' \emph{IEEE Trans. Commun.}, vol.~72, no.~10, pp. 6553--6568, Oct. 2024.

\bibitem{Rs2}
H.~Yang, A.~Zhang, Y.~Sun, J.~Li, and P.~Liu, ``Regularized sparse {B}ayesian learning based channel estimation for {RIS}-assisted wireless communication system,'' \emph{IEEE Commun. Lett.}, vol.~28, no.~6, pp. 1412--1416, Mar. 2024.

\bibitem{Rs1}
J.~Cao, J.~Du, M.~Han, J.~Liu, X.~Li, and D.~B. da~Costa, ``Efficient sparse {B}ayesian channel estimation for near-field ultra-scale massive {MIMO} systems,'' \emph{IEEE Wirel. Commun. Lett.}, vol.~12, no.~12, pp. 2133--2137, Dec 2023.

\bibitem{r27}
J.~Tian, Y.~Han, S.~Jin, X.~Li, J.~Zhang, and M.~Matthaiou, ``Near-field channel reconstruction in sensing {RIS}-assisted wireless communication systems,'' \emph{IEEE Trans. Wireless Commun.}, vol.~23, no.~9, pp. 12\,223--12\,238, Apr. 2024.

\bibitem{r28}
T.~Lin, X.~Yu, Y.~Zhu, and R.~Schober, ``Channel estimation for {IRS}-assisted millimeter-wave mimo systems: Sparsity-inspired approaches,'' \emph{IEEE Trans. Commun.}, vol.~70, no.~6, pp. 4078--4092, Apr. 2022.

\bibitem{r29}
Z.~Wei, D.~Miaomiao, and K.~Taejoon, ``{MMV}-based sequential {AoA} and {AoD} estimation for millimeter wave {MIMO} channels,'' \emph{IEEE Trans. Commun.}, vol.~70, no.~6, pp. 4063--4077, Apr. 2022.

\bibitem{r30}
X.~Zhou, X.~Liu, G.~Zhang, L.~Jia, X.~Wang, and Z.~Zhao, ``An iterative threshold algorithm of {Log-Sum} regularization for sparse problem,'' \emph{IEEE Trans. Circuits Syst. Video Technol.}, vol.~33, no.~9, pp. 4728--4740, Feb. 2023.

\bibitem{r33}
M.~V.~W. Zibetti, E.~S. Helou, and D.~R. Pipa, ``Accelerating overrelaxed and monotone fast iterative shrinkage-thresholding algorithms with line search for sparse reconstructions,'' \emph{IEEE Trans. Image Process.}, vol.~26, no.~7, pp. 3569--3578, Apr. 2017.

\bibitem{r34}
M.~V.~W. Zibetti, E.~S. Helou, R.~R. Regatte, and G.~T. Herman, ``Monotone {FISTA} with variable acceleration for compressed sensing magnetic resonance imaging,'' \emph{IEEE Trans. Comput. Imaging}, vol.~5, no.~1, pp. 109--119, Nov. 2019.

\bibitem{r35}
T.~Ahmed, X.~Zhang, and W.~U. Hassan, ``A higher-order propagator method for {2D-DOA }estimation in massive {MIMO} systems,'' \emph{IEEE Commun. Lett.}, vol.~24, no.~3, pp. 543--547, Dec. 2020.

\bibitem{r36}
Y.~Liu, S.~Zhang, F.~Gao, J.~Tang, and O.~A. Dobre, ``Cascaded channel estimation for {RIS} assisted mmwave {MIMO} transmissions,'' \emph{IEEE Wirel. Commun. Lett.}, vol.~10, no.~9, pp. 2065--2069, Jun. 2021.

\bibitem{r37}
E.~M. Abadi, A.~N. Hokmabadi, and S.~Gezici, ``Joint {RIS} phase profile design and power allocation for parameter estimation in presence of eavesdropping,'' \emph{IEEE Trans. Veh. Technol.}, vol.~73, no.~12, pp. 19\,186--19\,202, Aug. 2024.

\bibitem{r38}
M.~Varanasi, C.~Mullis, and A.~Kapur, ``On the limitation of linear {MMSE} detection,'' \emph{IEEE Trans. Inf. Theory}, vol.~52, no.~9, pp. 4282--4286, Aug. 2006.

\bibitem{r40}
D.~Ding, Y.~Zeng, and D.~Wang, ``Channel estimation for delay alignment modulation,'' in \emph{IEEE Wireless Commun. Netw. Conf. (WCNC)}, Dubai, United Arab Emirates, 2024, pp. 1--6.

\bibitem{r39}
K.~Dovelos, M.~Matthaiou, H.~Q. Ngo, and B.~Bellalta, ``Channel estimation and hybrid combining for wideband terahertz massive {MIMO} systems,'' \emph{IEEE J. Sel. Areas Commun.}, vol.~39, no.~6, pp. 1604--1620, Apr. 2021.

\end{thebibliography}

\begin{IEEEbiography}[{\includegraphics[width=1in,height=1.25in,clip,keepaspectratio]{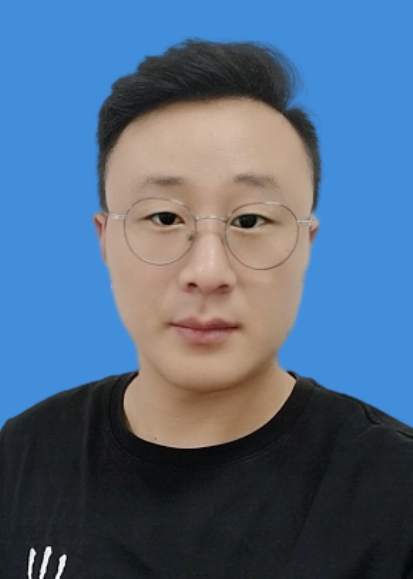}}] {Wenzhou Cao} received the B.S. and M.S. degrees in electronic and information engineering from the Zhongyuan University of Technology (ZUT), Zhengzhou, China, in 2018 and 2021, respectively. He is currently pursuing the Ph.D. degree in the School of Information and Communication Engineering, Beijing University of Posts and Telecommunications (BUPT), Beijing, China. His research interests include intelligent reflecting surface (IRS), channel estimation, and tensor decomposition.
\end{IEEEbiography}

\begin{IEEEbiography}[{\includegraphics[width=1in,height=1.25in,clip,keepaspectratio]{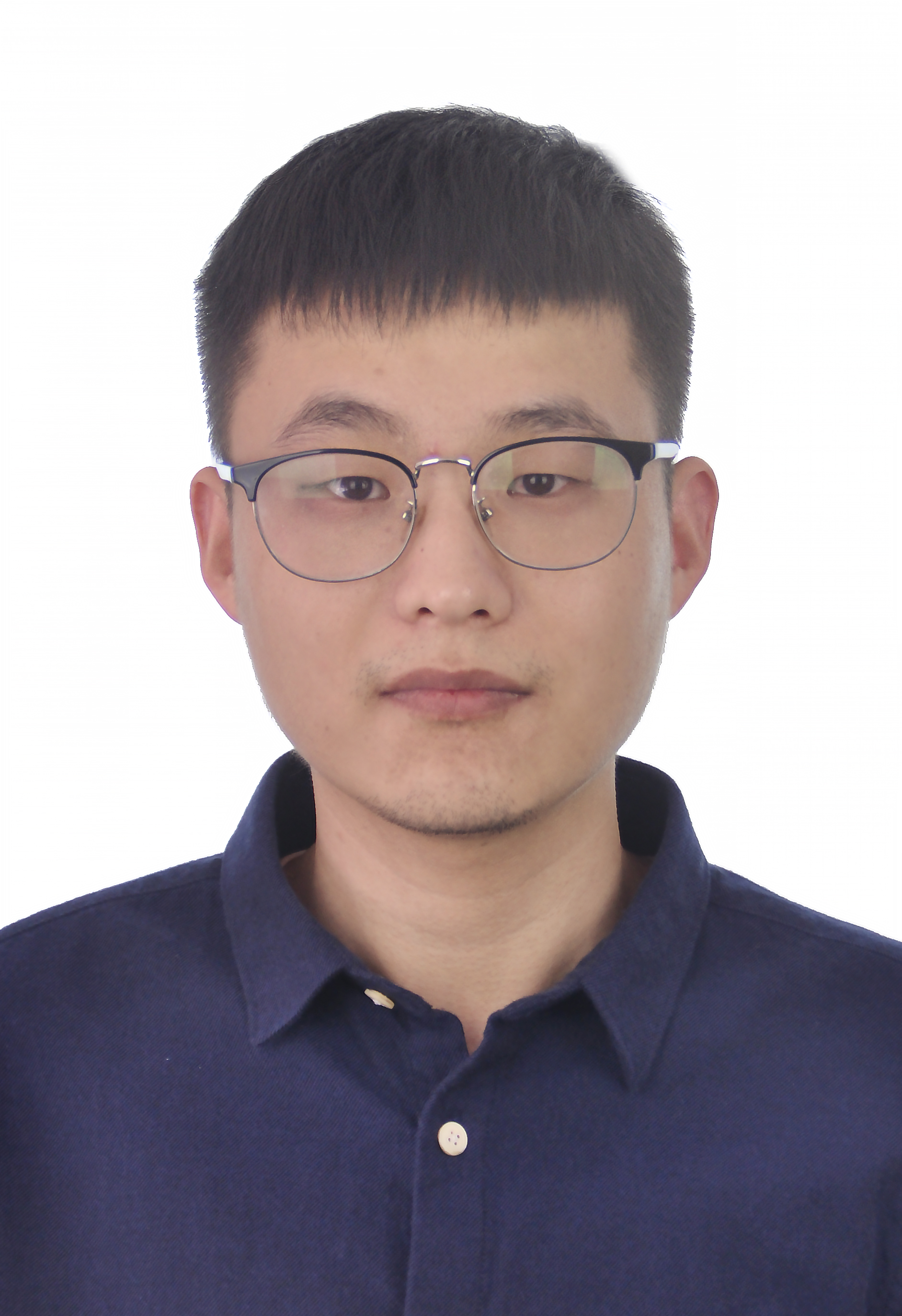}}]{Yashuai Cao} received the B.E. and Ph.D. degrees in communication engineering from Chongqing University of Posts and Telecommunications (CQUPT) and Beijing University of Posts and Telecommunications (BUPT), China, in 2017 and 2022, respectively. From 2022 to 2023, he was a lecturer in the Department of Electronics and Communication Engineering, North China Electric Power University (NCEPU), Baoding. From 2023 to 2025, he was a Postdoctoral Research Fellow with the Department of Electronic Engineering, Tsinghua University, Beijing, China. He is currently a Distinguished Associate Professor with the School of Intelligence Science and Technology, University of Science and Technology Beijing (USTB), Beijing, China. His research interests include Stacked Intelligent Metasurface, Environment-Aware Communications, and Channel Knowledge Map.
\end{IEEEbiography}

\begin{IEEEbiography}[{\includegraphics[width=1in,height=1.25in,clip,keepaspectratio]{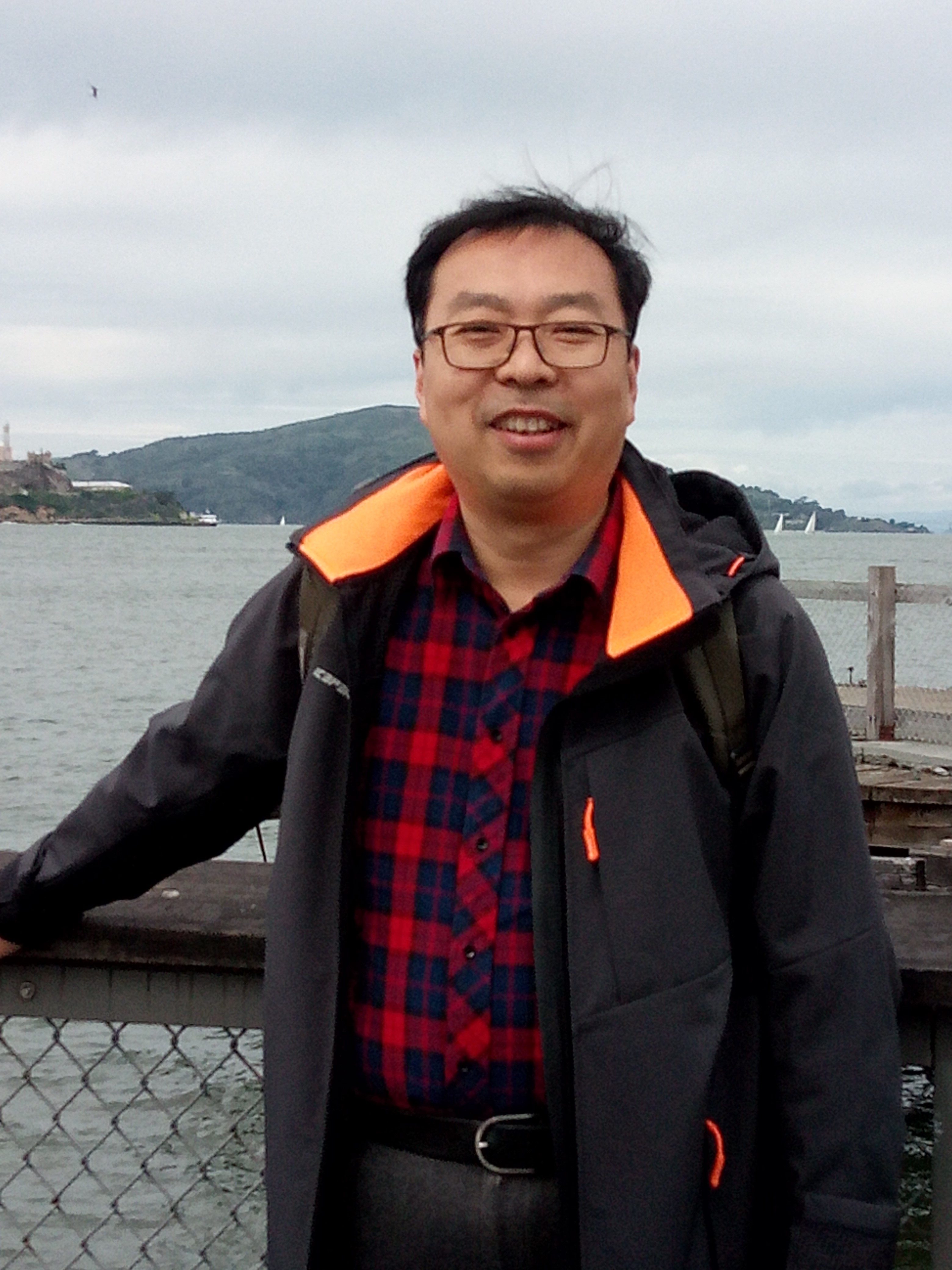}}]{Tiejun Lv} (Senior Member, IEEE) received the M.S. and Ph.D. degrees in electronic engineering from the University of Electronic Science and Technology of China (UESTC), Chengdu, China, in 1997 and 2000, respectively. From January 2001 to January 2003, he was a Post-Doctoral Fellow at Tsinghua University, Beijing, China. In 2005, he was promoted to a Full Professor at the School of Information and Communication Engineering, Beijing University of Posts and Telecommunications (BUPT). From September 2008 to March 2009, he was a Visiting Professor with the Department of Electrical Engineering, Stanford University, Stanford, CA, USA. He is currently the author of four books, one book chapter, and more than 170 published journal articles and 200 conference papers on the physical layer of wireless mobile communications. His current research interests include signal processing, communications theory, and networking. He was a recipient of the Program for New Century Excellent Talents in University Award from the Ministry of Education, China, in 2006. He received the Nature Science Award from the Ministry of Education of China for the hierarchical cooperative communication theory and technologies in 2015 and Shaanxi Higher Education Institutions Outstanding Scientific Research Achievement Award in 2025.
\end{IEEEbiography}

\begin{IEEEbiography}[{\includegraphics[width=1in,height=1.25in,clip,keepaspectratio]{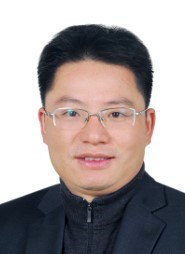}}]{Mugen Peng} received the Ph.D. degree in communication and information systems from the Beijing University of Posts and Telecommunications (BUPT), Beijing, China, in 2005. Afterward, he joined BUPT, where he has been the Dean of the School of Information and Communication Engineering since June 2020 and the Deputy Director of the State Key Laboratory of Networking and Switching Technology since October 2018. In 2014, he was also an Academic Visiting Fellow with Princeton University, USA. He has authored or coauthored over 150 refereed IEEE journal articles and over 250 conference proceeding papers. His main research areas include wireless communication theory, radio signal processing, cooperative communication, cloud communication, and the Internet of Things. He was a recipient of the 2018 Heinrich Hertz Prize Paper Award, the 2014 IEEE ComSoc AP Outstanding Young Researcher Award, and the Best Paper Award in the ICC 2022, ICCC 2020, IEEE WCNC 2015, and JCN 2016. He is currently on the Editorial/Associate Editorial Board Member of the IEEE Network, the IEEE Communications Magazine, the IEEE Internet of Things Journal, the IEEE Transactions on Vehicular Technology, the IEEE Transactions on Network Science and Engineering, the Intelligent and Converged Networks, and the Digital Communications and Networks (DCN).
\end{IEEEbiography}

\end{document}